\documentclass[aps,pra,11pt,amsmath,amssymb,tightenlines,superscriptaddress,notitlepage,nofootinbib,onecolumn]{revtex4-2}
\pdfoutput=1

\usepackage{graphicx}
\usepackage[dvipsnames,svgnames]{xcolor}
\usepackage{bm,bbm}%
\usepackage{braket}
\usepackage{amsthm,amsmath,amssymb,amsbsy}
\usepackage{enumerate}
\usepackage[shortlabels]{enumitem}
\usepackage{booktabs,multirow,moresize}
\usepackage{mathtools,mathrsfs,bbding}
\usepackage{microtype}

\usepackage{chngcntr}

\usepackage[T1]{fontenc}
\usepackage[utf8]{inputenc}

\definecolor{darkred}{rgb}{0.57,0,0.12}
\usepackage[colorlinks=true, linkcolor=MidnightBlue, urlcolor=MidnightBlue, citecolor=MidnightBlue]{hyperref}

\usepackage{cleveref}

\let\nc\newcommand

\let\mathscr\relax
\usepackage{newpxtext,newpxmath}
\usepackage[cal=cm]{mathalfa}

\hypersetup{
    bookmarksnumbered=true, 
    breaklinks=true,
    unicode=false, 
    pdfstartview={FitH}, 
    pdftitle={}, 
    pdfnewwindow=true, 
    colorlinks=true, 
    allcolors=MidnightBlue!70!black!70!TealBlue
}

\nc{\note}[1]{{\color{blue!90!black} #1}}
\nc{\noteb}[1]{{\color{red!80!black}\textbf{#1}}}

\DeclareMathOperator{\Tr}{Tr}

\DeclareMathOperator{\NN}{NN}
\DeclareMathOperator{\supp}{supp}

\DeclareMathOperator{\ran}{ran}

\DeclareMathOperator{\coll}{coll}

\let\Re\relax
\DeclareMathOperator{\Re}{Re}

\nc{\psic}{\psi^{c}}
\nc{\two}[1]{\underline{2^{d-#1}}}

\nc{\Hanc}{\mathcal{H}_{\text{anc}}}
\nc{\psianc}{\psi_{\text{anc}}}
\nc{\lampow}{\lambda^{1/d}}

\nc{\norm}[2]{\left\lVert#1\right\rVert_{#2}}
\nc{\proj}[1]{\ket{#1}\!\bra{#1}}
\nc{\pro}[1]{#1 #1^\dagger}
\nc{\lnorm}[2]{\left\lVert#1\right\rVert_{\ell_{#2}}}

\nc{\VV}{V^{\,\C}_\Sp}

\nc{\R}{\mathcal{R}}
\nc{\W}{\mathcal{W}}
\nc{\T}{\mathcal{T}}
\nc{\B}{\mathcal{B}}
\nc{\C}{\mathcal{C}}
\nc{\U}{\mathcal{U}}
\nc{\E}{\mathcal{E}}
\DeclareMathOperator*{\EE}{\mathbb{E}}
\nc{\K}{\mathcal{K}}
\nc{\V}{\mathcal{V}}
\nc{\X}{\mathcal{X}}
\nc{\F}{\mathcal{F}}
\nc{\Z}{\mathcal{Z}}
\nc{\G}{\mathcal{G}}
\nc{\D}{\mathcal{D}}

\nc{\Y}{\mathcal{Y}}

\nc{\M}{\mathcal{M}}
\nc{\N}{\mathcal{N}}
\nc{\I}{\mathcal{I}}
\nc{\Q}{\mathbb{Q}}
\nc{\RR}{\mathbb{R}}
\nc{\CC}{\mathbb{C}}

\nc{\HH}{\mathbb{H}}
\nc{\MM}{\smash{\mathbb{M}}}
\nc{\DD}{\mathbb{D}}
\renewcommand{\NN}{\mathbb{N}}

\nc{\J}{\mathcal{J}}
\nc{\frF}{\mathfrak{F}}

\newcommand{\univ}[1]{\mathfrak{U}_{#1}}
\newcommand{\univstar}[1]{\mathfrak{U}_{#1}\mkern-7mu\smash{\raisebox{2pt}{\*}}}
\newcommand{\univstars}[1]{\mathfrak{U}_{#1}\mkern-7mu\smash{\raisebox{1pt}{\scriptsize \*}}}

\nc{\Bures}{\mspace{1mu}}

\nc{\FF}{\mathbb{F}}

\nc{\CPTP}{{\mathrm{CPTP}}}
\nc{\CP}{{\mathrm{CP}}}
\nc{\CPTN}{{\mathrm{CPTN}}}

\nc{\ext}[1]{\operatorname{ext}\left(#1\right)}

\renewcommand{\*}{\textup{*}}
\nc{\<}{\left\langle}
\renewcommand{\>}{\right\rangle}

\let\bar\relax
\nc{\lset}{\left\{\vphantom{\big|}}
\nc{\bar}{\;\middle|\;}
\nc{\rset}{\vphantom{\big|}\right\}}

\nc{\ve}{\varepsilon}

\nc{\id}{\mathbbm{1}}
\nc{\idc}{\mathrm{id}}

\nc{\RHS}{(\mathrm{RHS})}

\theoremstyle{plain}

\newtheorem{theorem}{Theorem}

\newtheorem{proposition}[theorem]{Proposition}
\newtheorem{corollary}[theorem]{Corollary}
\newtheorem{lemma}[theorem]{Lemma}

\theoremstyle{definition}

\newtheorem*{remark}{Remark}

\let\oldproofname\proofname
\renewcommand{\proofname}{\rm\bf{\oldproofname}}

\makeatletter
\renewenvironment{proof}[1][\proofname]{\par
\pushQED{\qed}%
\normalfont \topsep6\p@\@plus6\p@\relax
\trivlist
\item\relax
{\bfseries  
#1\@addpunct{.}}\hspace\labelsep\ignorespaces 
}{%
\popQED\endtrivlist\@endpefalse
}
\makeatother

\nc{\ketbra}[2]{\ket{#1}\!\bra{#2}}

\nc{\prob}{\mathrm{prob}}

\makeatletter
\let\save@mathaccent\mathaccent
\newcommand*\if@single[3]{%
  \setbox0\hbox{${\mathaccent"0362{#1}}^H$}%
  \setbox2\hbox{${\mathaccent"0362{\kern0pt#1}}^H$}%
  \ifdim\ht0=\ht2 #3\else #2\fi
  }
\newcommand*\rel@kern[1]{\kern#1\dimexpr\macc@kerna}
\newcommand*\widebar[1]{\wide@bar{#1}{0}}
\newcommand*\wide@bar[2]{\if@single{#1}{\wide@bar@{#1}{#2}{1}}{\wide@bar@{#1}{#2}{2}}}
\newcommand*\wide@bar@[3]{%
  \begingroup
  \def\mathaccent##1##2{%
    \let\mathaccent\save@mathaccent
    \if#32 \let\macc@nucleus\first@char \fi
    \setbox\z@\hbox{$\macc@style{\macc@nucleus}_{}$}%
    \setbox\tw@\hbox{$\macc@style{\macc@nucleus}{}_{}$}%
    \dimen@\wd\tw@
    \advance\dimen@-\wd\z@
    \divide\dimen@ 3
    \@tempdima\wd\tw@
    \advance\@tempdima-\scriptspace
    \divide\@tempdima 10
    \advance\dimen@-\@tempdima
    \ifdim\dimen@>\z@ \dimen@0pt\fi
    \rel@kern{0.6}\kern-\dimen@
    \if#31
      \overline{\rel@kern{-0.6}\kern\dimen@\macc@nucleus\rel@kern{0.4}\kern\dimen@}%
      \advance\dimen@0.4\dimexpr\macc@kerna
      \let\final@kern#2%
      \ifdim\dimen@<\z@ \let\final@kern1\fi
      \if\final@kern1 \kern-\dimen@\fi
    \else
      \overline{\rel@kern{-0.6}\kern\dimen@#1}%
    \fi
  }%
  \macc@depth\@ne
  \let\math@bgroup\@empty \let\math@egroup\macc@set@skewchar
  \mathsurround\z@ \frozen@everymath{\mathgroup\macc@group\relax}%
  \macc@set@skewchar\relax
  \let\mathaccentV\macc@nested@a
  \if#31
    \macc@nested@a\relax111{#1}%
  \else
    \def\gobble@till@marker##1\endmarker{}%
    \futurelet\first@char\gobble@till@marker#1\endmarker
    \ifcat\noexpand\first@char A\else
      \def\first@char{}%
    \fi
    \macc@nested@a\relax111{\first@char}%
  \fi
  \endgroup
}
\makeatother

\nc{\wt}{\widetilde}
\renewcommand{\ol}{\widebar}

\nc{\rhos}{\rho'}

\let\epsilon\varepsilon

\usepackage{scalerel,stackengine}

\newcommand{\deff}[1]{\textbf{\emph{#1}}}

\nc{\ba}{\begin{equation}\begin{aligned}}
\nc{\ea}{\end{aligned}\end{equation}}

\let\textleq\relax

\let\texteq\relax

\newcommand{\texteq}[1]{\stackrel{\mathclap{\mbox{\scriptsize #1}}}{=}}
\newcommand{\textleq}[1]{\stackrel{\mathclap{\mbox{\scriptsize #1}}}{\leq}}

\usepackage[most,breakable]{tcolorbox}
\renewenvironment{boxed}[1][white]%
  {\expandafter\ifstrequal\expandafter{#1}{filled}{\begin{tcolorbox}[colback=MidnightBlue!70!black!70!TealBlue!2!white,colframe=MidnightBlue!70!black!70!TealBlue!30!white,breakable=false,enhanced,left=5.75pt,right=5.75pt,grow sidewards by=10pt]}{\begin{tcolorbox}[colback=white,colframe=gray!15,breakable,enhanced,left=5.75pt,right=5.75pt,grow sidewards by=10pt]}}%
  {\end{tcolorbox}}
  \newenvironment{tboxed}[2][white]%
  {\expandafter\ifstrequal\expandafter{#1}{filled}{\begin{tcolorbox}[colback=MidnightBlue!70!black!70!TealBlue!4!white,colframe=MidnightBlue!70!black!70!TealBlue!20!white,breakable=false,enhanced,left=5.75pt,right=5.75pt,grow sidewards by=10pt,title={\color{black}\textbf{#2}}]}{\begin{tcolorbox}[colback=white,colframe=gray!15,enhanced,left=5.75pt,right=5.75pt,grow sidewards by=10pt,title={\color{black}\textbf{#2}}]}}%
  {\end{tcolorbox}}

\nc{\rrho}{\widetilde{\rho}}

\nc{\abs}[1]{\left|#1\right|}

\usepackage{stackengine}

\renewcommand{\thesubsection}{\arabic{subsection}}

\usepackage{tikz-cd}

\makeatletter
\def\fnum@figure{\textbf{Figure}\nobreakspace\textbf{\thefigure}}
\renewcommand{\p@subsection}{\thesection.}
\renewcommand{\p@subsubsection}{\thesection.\thesubsection.}
\makeatother

\DeclareMathOperator{\opt}{opt}

\DeclareMathOperator{\err}{err}

\usepackage{xstring}
\usepackage{xparse}

\NewDocumentCommand{\s}{m m o}{%
 \lowercase{\def\sdist{#2}}%
 ^{#1,\mspace{2mu}%
 \IfEqCase{\sdist}{%
  {t}{%
    T%
  }%
  {p}{%
    P%
  }%
  {m}{%
    \MM%
  }%
 }%
 \IfNoValueTF{#3}{}{%
    \lowercase{\def\sarr{#3}}%
    ,%
    \IfEqCase{\sarr}{%
      {up}{ \mspace{.5mu}\smash\uparrow }%
      {uparrow}{ \mspace{.5mu}\smash\uparrow }%
      {down}{ \mspace{1mu}\smash\downarrow }%
      {downarrow}{ \mspace{1mu}\smash\downarrow }%
    }%
  }%
 }
 \IfNoValueTF{#3}{}{%
  \mathchoice{\vphantom{^{\uparrow}}}{}{}{}%
 }%
}%
\NewDocumentCommand{\Ds}{m m m}{%
  D_{#1}\s{#2}{#3}%
}

\NewDocumentCommand{\Hs}{m m m O{up}}{%
  H_{#1}\s{#2}{#3}[#4]%
}
\NewDocumentCommand{\Hmin}{m m O{up}}{%
  H_{\min}\s{#1}{#2}[#3]%
}
\makeatletter

\def\@sect@ltx#1#2#3#4#5#6[#7]#8{%
    \@ifnum{#2>\c@secnumdepth}{%
        \def\H@svsec{\phantomsection}%
        \let\@svsec\@empty
    }{%
        \H@refstepcounter{#1}%
        \def\H@svsec{%
            \phantomsection
        }%
        \protected@edef\@svsec{{#1}}%
        \@ifundefined{@#1cntformat}{%
            \prepdef\@svsec\@seccntformat
        }{%
            \expandafter\prepdef
            \expandafter\@svsec
            \csname @#1cntformat\endcsname
        }%
    }%
    \@tempskipa #5\relax
    \@ifdim{\@tempskipa>\z@}{%
        \begingroup
        \interlinepenalty \@M
        #6{%
            \@ifundefined{@hangfrom@#1}{\@hang@from}{\csname @hangfrom@#1\endcsname}%
            {\hskip#3\relax\H@svsec}{\@svsec}{#8}%
        }%
        \@@par
        \endgroup
        \@ifundefined{#1mark}{\@gobble}{\csname #1mark\endcsname}{#7}%
        \addcontentsline{toc}{#1}{%
            \@ifnum{#2>\c@secnumdepth}{%
                \protect\numberline{}%
            }{%
                \protect\numberline{\csname the#1\endcsname}%
            }%
            #7}
    }{%
        \def\@svsechd{%
            #6{%
                \@ifundefined{@runin@to@#1}{\@runin@to}{\csname @runin@to@#1\endcsname}%
                {\hskip#3\relax\H@svsec}{\@svsec}{#8}%
            }%
            \@ifundefined{#1mark}{\@gobble}{\csname #1mark\endcsname}{#7}%
            \addcontentsline{toc}{#1}{%
                \@ifnum{#2>\c@secnumdepth}{%
                    \protect\numberline{}%
                }{%
                    \protect\numberline{\csname the#1\endcsname}%
                }%
                #8}%
        }%
    }%
    \@xsect{#5}}%

\makeatother

\let\oldaddcontentsline\addcontentsline
\newcommand{\excludetoc}{\renewcommand{\addcontentsline}[3]{}}
\newcommand{\includetoc}{\let\addcontentsline\oldaddcontentsline}

\newcommand{\row}{L_1(\ell_2^{R})}
\newcommand{\col}{L_1(\ell_2^{C})}
\newcommand{\rowcol}{L_1(\ell_2^{RC})}

\usepackage{mathcomp}

\usepackage{pict2e}

\DeclareRobustCommand{\lrighttriangle}{%
  \begingroup
  \setlength{\unitlength}{.9ex}%
  \begin{picture}(1,1)
  \polyline(1,0)(0,0)(0,1)(1,0)(.5,0)
  \end{picture}%
  \endgroup
}
\DeclareRobustCommand{\rrighttriangle}{%
  \begingroup
  \setlength{\unitlength}{.9ex}%
  \begin{picture}(1,1)
  \polyline(1,0)(0,0)(1,1)(1,0)(.5,0)
  \end{picture}%
  \endgroup
}
\newcommand{\hq}[1]{\overset{\lrighttriangle}{#1}}
\newcommand{\hqr}[1]{\overset{\rrighttriangle}{#1}}
\newcommand{\fqr}[1]{\overset{\square}{#1}}

\begin{document}

\title{Sharp one-shot bounds and optimal exponents\texorpdfstring{\\}{ }of universal randomness extraction}

\author{Bartosz Regula}
\email{bartosz.regula@gmail.com}
\affiliation{Mathematical Quantum Information RIKEN Hakubi Research Team, RIKEN Pioneering Research Institute (PRI) and RIKEN Center for Quantum Computing (RQC), Wako, Saitama, Japan}

\author{Hao-Chung Cheng}
\email{haochung.ch@gmail.com}
\affiliation{Department of Electrical Engineering and Graduate Institute of Communication Engineering, National Taiwan University, Taipei 106, Taiwan}
\affiliation{Physics and Mathematics Divisions, National Center for Theoretical Sciences, Taipei 10617, Taiwan}
\affiliation{Hon Hai (Foxconn) Quantum Computing Center, New Taipei City 236, Taiwan}

\begin{abstract}
We introduce techniques for the study of randomness extraction against quantum side information (privacy amplification) by relating this problem to variants of hypothesis testing tasks with quadratic constraints. 
Building also on the recently introduced measured smooth entropy approach of [Regula and Tomamichel, \href{https://arxiv.org/abs/2603.04493}{arXiv:2603.04493}], we precisely determine the optimal scaling of the error of quantum privacy amplification in several regimes.
Our results apply to the standard operational security criterion based on trace distance and focus in particular on universal classes of hash functions.  
On the technical side, our investigation reveals connections with operator space theory methods used in non-commutative Khintchine inequalities. 
As one application, we give a one-shot converse bound that shows the leftover hash lemma of [\href{https://arxiv.org/abs/2603.04493}{arXiv:2603.04493}] to be tight up to a constant factor for all strongly 4-universal hash functions. This allows us to evaluate the asymptotic i.i.d.\ error exponent of universal randomness extraction for all rates, matching the bound of~[Dupuis, \href{https://ieeexplore.ieee.org/document/10232924}{IEEE T-IT (2023)}] and improving also earlier classical asymptotic results of [Hayashi, \href{https://arxiv.org/abs/1010.1358}{IEEE T-IT (2013)}].
We furthermore give new one-shot achievability and converse bounds on the error in the strong converse regime, which we show to match up to a constant factor. This establishes the exact strong converse exponent of quantum privacy amplification under 2-universal hash functions, complementing the converse bound of [Shen, Gao, and Cheng, \href{https://arxiv.org/abs/2202.10263}{arXiv:2202.10263}].
\end{abstract}

\date{\today}

\maketitle

\tableofcontents


\section{Introduction}

Applications to quantum key distribution~\cite{renner_2005,tomamichel_2011} and quantum information processing~\cite{renes_2010,renes_2018-1,dupuis_2010} motivate the study of randomness extraction against quantum side information (privacy amplification), that is, the task of converting imperfect sources of randomness into perfectly uniform randomness in a way that ensures security from any quantum adversary.  
Recent developments showed that known achievability results for this task can be substantially improved by reconsidering the non-commutative extensions of entropic quantities used for the study of quantum side information, and in particular the choice of the most suitable notion of smooth entropies~\cite{regula_2026,tan_2026}. 
This prompts us to investigate whether the achievability results could be improved further or if they are in fact optimal, in the sense of exactly determining the scaling of the best achievable error. It is, however, not clear how to tackle this question with existing tools, and even precisely defining what optimality should mean here may not be immediate.

An important aspect of privacy amplification is for it to be realised \emph{universally}, in the sense of not depending on the specific source distribution, which is also key to ensuring the security from adversaries who may be correlated with the source. This is enabled through suitable assumptions on the mathematical properties of the hash functions, often assumed to be taken from so-called $k$-universal families whose collision probabilities are constrained or whose mutual independence can be guaranteed~\cite{carter_1979,wegman_1981}. 
However, this also leads to difficulties in the analysis of the task, as the optimal behaviour of operational figures of merit such as the trace distance can depend on the properties of the hash functions~\cite{hayashi_2013}. 
While some aspects of quantum privacy amplification under trace distance are indeed fully characterised, including its optimal second-order asymptotic expansion~\cite{shen_2024,regula_2026}, and a number of bounds on the error scaling (in the sense of i.i.d.\ error and strong converse exponents) have been shown~\cite{dupuis_2023,shen_2022,regula_2026}, the optimal scaling of the error under trace distance has been established only in the classical case and only under a restricted type of universal hash functions~\cite{hayashi_2013,hayashi_2016-3}. Instead, results for quantum side information are often established for other measures of distance, such as the purified distance~\cite{tomamichel_2013,li_2023,li_2024,rubboli_2026} or other entropic measures~\cite{hayashi_2015,cheng_2025-1,li_2026-1,li_2026}.   
Since it is the trace distance that defines the operationally relevant figure of merit in quantum privacy amplification for applications such as quantum cryptography~\cite{portmann_2022,ferradini_2025}, determining the optimal error scaling for this distance is thus an important question that has remained open.

Here we will introduce a framework for the analysis of quantum privacy amplification that helps shed light on its structure through the lens of quantum hypothesis testing.
The idea that underlies and connects the methods is the incorporation of quadratic constraints into hypothesis testing problems.
Specifically, binary quantum hypothesis testing is concerned with discriminating between two states, $\rho$ and $\sigma$, using a POVM element $M$ that defines a two-outcome measurement $( M, \id - M )$ with corresponding type I and type II error probabilities being, respectively, $\Tr (\id - M)\rho$ and $\Tr M \sigma$. In asymmetric hypothesis testing, one is typically concerned with optimising one of the two errors over all measurements $M$, subject to constraints on the other; in symmetric hypothesis testing, one instead aims to minimize the total (or average) error as
\begin{equation}\begin{aligned}
  \err(\rho\|\sigma) = \inf \lset \Tr (\id - M)\rho + \Tr M \sigma \bar 0 \leq M \leq \id \rset.
\end{aligned}\end{equation}
If one were to quantify the errors here not with the expected value of the corresponding measurement operator, but rather of its square --- similar, for example, to figures of merit in mean square error estimation --- we would instead obtain the quadratic optimisation problem
\begin{equation}\begin{aligned}
  \fqr\err(\rho\|\sigma) = \inf \lset \Tr (\id - M)^2\rho + \Tr M^2 \sigma \bar 0 \leq M \leq \id \rset.
\end{aligned}\end{equation}
The optimal solution $M$ here corresponds to a variant of the so-called pretty good measurement~\cite{hausladen_1994,barnum_2002} (see details in Section~\ref{sec:strong-converse}), which has found a number of applications, notably as the foundation of many achievability proofs in quantum information~\cite{hayashi_2003,beigi_2014,cheng_2023-1,beigi_2024,cheng_2025}. 
Approaches based on the pretty good measurement have, however, found some apparent limitations: for example, while they can give converse bounds on the strong converse exponent of quantum privacy amplification, it remained uncertain whether such bounds are in fact achievable and whether the techniques can be used to analyse the behaviour of error exponents~\cite{shen_2022}. 
Our contribution here starts with the realisation that a more suitable type of optimisation problems can be built from \emph{half}-quadratic hypothesis-testing--type problems of the~form
\begin{equation}\begin{aligned}\label{eq:hqr_intro}
  \hqr\err(\rho\|\sigma) = \inf \lset \Tr (\id - M)\rho + \Tr M^2 \sigma \bar 0 \leq M \leq \id \rset.
\end{aligned}\end{equation}
The intuition here is that, while the quadratic constraint can help access quadratic functions of the involved state, the linear part more closely connects with linear functions like the trace distance, allowing us to control them more tightly. Connection with quantities relevant to the study of randomness extraction can be seen by considering a Lagrange dual of the optimisation in~\eqref{eq:hqr_intro}:
\begin{equation}\begin{aligned}\label{eq:hqr_intro_dual}
  1 - \hqr\err(\rho\|\sigma) = \inf_{\ve \in [0,1]} \left[ \ve + \frac14 \exp\!\Big( \Ds{2}{\ve}{M}(\rho\|\sigma)\Big) \right],
\end{aligned}\end{equation}
where $\Ds{2}{\ve}{M}$ is the measured smooth collision divergence, a variant of smooth R\'enyi relative entropy recently used by~\cite{regula_2026} as a key tool in deriving improved bounds on quantum randomness extraction. 
The form of Equation~\eqref{eq:hqr_intro_dual} already hints at two connections: one, it resembles the form that many one-shot achievability results take, including the leftover hash lemma that underpins privacy amplification~\cite{bennett_1995,tomamichel_2011,regula_2026}; two, interpreting $\ve$ as trace distance between the state $\rho$ and a smoothing operator that approximates it, the expression can be understood as a generalised interpolation functional between $L_1$- and $L_2$-type spaces, and such quantities find prominent use in the analysis of asymptotics of random variables~\cite{johnson_1989,hitczenko_1994}.
Building on these two connections, our approach will allow us to tightly characterise universal quantum privacy amplification.

\subsection{Summary of results}

We first show that the recent tightened variant of the leftover hash lemma by~\cite{regula_2026} can be understood precisely as a half-quadratic hypothesis testing expression: namely, the result bounds the expected trace distance from uniform and secure randomness as
\begin{equation}\begin{aligned}\label{eq:lhl_intro}
  \EE_f \frac12 \norm{\rho_{ZE}^f - \frac{\id_Z}{|\Z|} \otimes \rho_E}{1} \leq \hqr\pi_{|\Z|/4}(X|E)_\rho
\end{aligned}\end{equation}
where $\hqr\pi_{|\Z|/4}$ corresponds to the best discrimination probability, $\Tr M \rho_{XE}$, optimised over measurements $M$ subject to the quadratic constraint that $\Tr M^2 (\id_X \otimes \sigma_E) \leq \frac{|\Z|}{4}$ for all states $\sigma_E$. 
Here, $\rho_{ZE}^{\smash{f}}$ denotes the classical--quantum state $\rho_{XE}$ after the application of a hash function $f\colon \X \to \Z$ randomly chosen from a 2*-universal family of hashes, an assumption which only constrains the pairwise collision probabilities of the hash functions (see detailed definitions in Section~\ref{sec:prelim}).
This form already allows for a simplified derivation of some results of~\cite{regula_2026} such as bounds in terms of R\'enyi entropies, recovering leading bounds on error exponents~\cite{dupuis_2023}. 

We are then interested in the question of how tight the achievability result of Eq.~\eqref{eq:lhl_intro} is. 
While privacy amplification is commonly implemented using variants of 2-universal hash functions, 2-universality is in general not enough to establish universal lower bounds, and the hash families need additional assumptions that ensure sufficient randomisation. 
Here we show in particular that a tight one-shot converse to~\eqref{eq:lhl_intro} holds for all families of 4*-universal hash functions:
\begin{equation}\begin{aligned}\label{eq:lhl_intro_converse}
  \EE_f \frac12 \norm{\rho_{ZE}^f - \frac{\id_Z}{|\Z|} \otimes \rho_E}{1} \geq \frac13 \,\hqr\pi_{|\Z|/4}(X|E)_\rho.
\end{aligned}\end{equation}
Our assumption of 4*-universality here constrains the four-wise collision probabilities of the hash functions, and it is much less demanding than the assumptions previously used to argue converse bounds for randomness extraction against classical side information~\cite{hayashi_2013,hayashi_2016-3}.
The converse effectively proves that the scaling of the error indicated by the leftover hash lemma~\cite{regula_2026} is the tightest possible, since any 4*-universal family matches it tightly for any quantum state $\rho_{XE}$. As a consequence, it allows us to evaluate the optimal asymptotic error exponent of quantum privacy amplification under all  4*-universal hash functions: employing results in composite hypothesis testing against product states~\cite{hayashi_2016-1} we can evaluate the asymptotics of~\eqref{eq:lhl_intro_converse} for i.i.d.\ sources as
\begin{equation}\begin{aligned}
   \lim_{n\to\infty} - \frac1n \log  \left( \EE_{f_n} \frac12 \norm{ (\rho^{\otimes n}_{XE})^{f_n}_{Z_n E_n} - \frac{\id_{Z_n}}{|\Z_n|} \otimes \rho_E^{\otimes n} }{1} \right) = \sup_{\alpha \in (1,2]} \frac{\alpha-1}{\alpha} \left( \wt{H}^{\uparrow}_{\alpha}(X|E)_\rho - R \right)
\end{aligned}\end{equation}
where $|\Z_n| = \exp(nR)$ for any chosen rate $R > 0$, and $\wt{H}^{\uparrow}_{\alpha}$ denotes the sandwiched R\'enyi conditional entropy~\cite{muller-lennert_2013}. This proves the optimality of the  bound first shown by Dupuis~\cite{dupuis_2023} as the exact exponent of universal quantum privacy amplification under trace distance. In the classical case, our finding improves on the earlier asymptotic results of Hayashi and Watanabe~\cite{hayashi_2013,hayashi_2016-3}, which --- in addition to the more restrictive assumptions on the hash functions --- were only tight for sufficiently large rates $R$. 

Our approach to prove~\eqref{eq:lhl_intro_converse} is inspired by connections with operator space theory. 
As it turns out, the measured R\'enyi entropies that underlie the achievability results of~\cite{regula_2026} are in fact closely connected to fundamental operator norms used in the analysis of non-commutative random variables. In particular, the collision entropy $H_{2}^{\smash{\MM,\uparrow}}(X|E)_{\rho}$
that appears in~\cite{regula_2026}'s formulation of the quantum leftover hash lemma is equivalent to the same $L_1(\ell_2)$ row--column norm that governs non-commutative Khintchine inequalities~\cite{lust-piquard_1991,pisier_1997,junge_2008,pisier_2009}, and the related half-quadratic optimisation problem $\smash{\hqr\pi_{k}(X|E)_\rho}$ is a real interpolation functional built from this norm. 
To derive the tight one-shot result in~\eqref{eq:lhl_intro_converse}, we then adapt and extend the proof approach of Haagerup and Musat~\cite{haagerup_2007} that was first used to study optimal constants in operator Khintchine inequalities, and which we show to naturally connect with our half-quadratic hypothesis testing techniques.

In the opposite regime of strong converse exponents, one instead studies how fast the error converges to 1 when the demanded rate of randomness extraction $R$ is too large; this is conveniently expressed in terms of the asymptotics of the quantity $\err\!\left( \rho_{ZE}^{\smash{f}} \middle\| \frac{\id_Z}{|\Z|} \otimes \rho_E \right)$, since $\err(A\|B) = \frac12(\Tr A + \Tr B) - \frac12 \norm{A-B}{1}$ for any positive semidefinite operators. 
We first show a new achievability result~as
\begin{equation}\begin{aligned}
\EE_f \, \err\!\left( \rho_{ZE}^f  \middle\| \frac{\id_Z}{|\Z|} \otimes \rho_E \right) \geq \frac12  \err\!\left( \rho_{XE} \middle\|  \frac{\id_X}{|\Z|} \otimes \rho_E \right).
\end{aligned}\end{equation}
It uses the same assumptions as the leftover hash lemma of~\cite{regula_2026}, but does not follow from it directly: our approach combines the results of~\cite{regula_2026} with a careful characterisation of the properties of the half-quadratic error optimisation in~\eqref{eq:hqr_intro}. Adapting also the pretty good measurement technique as used in~\cite{shen_2022}, we give a matching one-shot converse bound
\begin{equation}\begin{aligned}
\EE_f \,\err\!\left( \rho_{ZE}^f  \middle\| \frac{\id_Z}{|\Z|} \otimes \rho_E \right) \leq 4 \err\!\left( \rho_{XE}  \middle\|  \frac{\id_X}{|\Z|} \otimes \rho_E \right),
\end{aligned}\end{equation}
which applies to any 2-universal family of hash functions. Combining the two, this gives an exact expression for the strong converse exponent of quantum privacy amplification under 2-universal hash functions:
\begin{equation}\begin{aligned}
\lim_{n\to\infty} - \frac1n \log \left( 1 -\EE_{f_n} \frac12 \norm{ (\rho^{\otimes n}_{XE})^{f_n}_{Z_n E_n} - \frac{\id_{Z_n}}{|\Z_n|} \otimes \rho_E^{\otimes n} }{1}\right) = \sup_{\alpha \in (0,1)} (\alpha-1) \left( \ol{H}^{\downarrow}_{\alpha}(X|E)_\rho - R \right),
\end{aligned}\end{equation}
which matches the converse bound of~\cite{shen_2022}. 
The entropic quantities that appear here are the Petz--R\'enyi conditional entropies $\ol{H}^{\downarrow}_{\alpha}$, rather than the sandwiched ones $\wt{H}^{\uparrow}_{\alpha}$ that govern the error exponents.

Our approach thus overcomes the difficulties in characterising the trace distance as an error metric for quantum privacy amplification and allows us to optimally characterise its behaviour under universal hash functions. Our findings indicate in particular that the constraints stated in terms of measured R\'enyi entropies $H_{\alpha}^{\MM}$ and real interpolation functionals based on such entropies are sharp. 
A notable aspect of the results is that the tightness of such bounds can be shown already at the one-shot level and up to constant factors, contrasting with many results in quantum information theory that are typically only able to establish optimality in an asymptotic sense.
This thus not only proves the optimality of earlier bounds on asymptotic error and strong converse exponents~\cite{dupuis_2023,shen_2022}, but it shows that already the smooth entropic one-shot bounds of~\cite{regula_2026} very tightly govern the optimal error scaling in a broad range of regimes. 
Coupled with earlier optimality results that apply in complementary contexts such as second-order analysis~\cite{regula_2026}, this gives further strong evidence for the suitability of this notion of smoothing for the analysis of quantum information processing tasks.


\section{Preliminaries}\label{sec:prelim}


\subsection{Quantum states, divergences, and entropies}

We will state our results for quantum states on finite-dimensional Hilbert spaces. We use lowercase Latin letters to refer to classical probability distributions on some finite alphabet $\X$, equivalently understood as diagonal quantum states in some orthonormal basis $\{\ket{x}\}_x$, i.e.\ $p_X = \sum_x p_X(x) \proj{x}$. A state $\rho_{XE}$ is called \emph{classical--quantum} (CQ) if $\rho_{XE} = \sum_x p_X(x) \proj{x} \otimes \ol{\rho}_{E,x}$ for some probability distribution $p_X$ and a collection of quantum states $\{\ol{\rho}_{E,x}\}_x$ on the system $E$; for simplicity, we will write this as $\rho_{XE} = \sum_x \proj{x} \otimes \rho_{E,x}$ where each operator $\rho_{E,x} = p_X(x) \ol{\rho}_{E,x}$ is understood as a subnormalised quantum state. 

To measure distance between quantum states, we will employ the trace distance (total variation distance) $\frac12 \| \rho - \rho' \|_1$, where $\norm{X}{1} = \Tr \sqrt{X^\dagger X}$ is the Schatten 1-norm. In order to account for subnormalised states, it will often be convenient to use the \emph{generalised trace distance} $\norm{\rho - \rho'}{+}$~\cite{tomamichel_2016}, where
\begin{equation}\begin{aligned}
  \norm{X}{+} &\coloneqq \frac12 \big|\!\Tr X\big| + \frac12 \|X\|_1 = \max_{0 \leq M \leq \id} \big|\! \Tr M X \big|.
\end{aligned}\end{equation}
Note that when $\Tr \rho' = \Tr \rho$, we have $\norm{\rho - \rho'}{+} = \frac12 \norm{\rho - \rho'}{1} = \Tr (\rho - \rho')_+$, where $X_+$ denotes the positive part of a self-adjoint operator. 
We will also use the notation $A \wedge B \coloneqq A - (A-B)_+$ to denote the non-commutative minimal between two positive semidefinite operators; it is naturally connected to the trace distance as $\Tr [A \wedge B] = \frac12\Tr(A+B) - \frac12 \norm{A-B}{1}$. 

For classical distributions, R\'enyi relative entropies are defined as
\begin{equation}\begin{aligned}
  D_\alpha(p\|q) \coloneqq \frac{1}{\alpha-1} \log \sum_x p(x)^{\alpha} \, q(x)^{1-\alpha}
\end{aligned}\end{equation}
with $\alpha \in (0,1)\cup(1,\infty)$; the extremal quantities $D_0$ and $D_\infty$ are defined as the corresponding limits in $\alpha$. 
Here we take the functions $\log$ and $\exp$ to the same, but otherwise arbitrary, base. 
From any R\'enyi relative entropy and $\ve \in [0,1]$, one constructs the \emph{smooth} R\'enyi divergence $D_\alpha^\ve$~\cite{renner_2004,renner_2005-2}~as
\begin{equation}\begin{aligned}
 D_\alpha^\ve(p\|q) \coloneqq \opt \lset D_{\alpha}(p'\|q) \bar p' \geq 0,\; \norm{p - p'}{+} \leq \ve,\; \Tr p' \leq 1 \rset,
\end{aligned}\end{equation}
where `$\opt$' stands for $\inf$ when $\alpha >1$ and $\opt = \sup$ for $\alpha < 1$.
The optimisation here can, without loss of generality, be restricted to sub-distributions $p'$ satisfying the element-wise inequality $p' \leq p$ (cf.~\cite{yang_2019,gour_2026}), making $p - p' \geq 0$ take on the intuitive interpretation as a probability mass discarded from $p$.

In the quantum case, many inequivalent variants of R\'enyi relative entropies and smoothing notions can be defined; we refer to~\cite{rubboli_review} for a comprehensive discussion. 
The main types of R\'enyi divergences are the Petz--R\'enyi relative entropies $\ol{D}_\alpha$~\cite{petz_1986}, the sandwiched R\'enyi relative entropies $\wt{D}_\alpha$~\cite{muller-lennert_2013,wilde_2014}, and the measured R\'enyi divergences $D_\alpha^{\MM}$~\cite{fuchs_1996,matsumoto_2014,berta_2017-1}, given respectively~by
\begin{equation}\begin{aligned}\label{eq:quantum_Renyi_def}
\ol{D}_\alpha (\rho \| \sigma) &\coloneqq \frac{1}{\alpha-1} \log \Tr \left( \rho^\alpha \sigma^{1-\alpha}\right),  \\
\wt{D}_\alpha (\rho \| \sigma) &\coloneqq \frac{1}{\alpha-1} \log \Tr\! \left[ \left(\sigma^{\frac{1-\alpha}{2\alpha}} \rho \sigma^{\frac{1-\alpha}{2\alpha}}\right)^\alpha \right],\\
D_\alpha^{\MM}(\rho\|\sigma) &\coloneqq \sup_{\M \in \MM} D_{\alpha}\! \left( \M(\rho) \middle\| \M(\sigma) \right)
\end{aligned}\end{equation}
whenever $\supp \rho \subseteq \supp \sigma$, and otherwise as the limits $\delta \to 0^+$ of the divergences evaluated between $\rho$ and $\sigma + \delta \id$. Here, $\MM$ denotes the set of all measurement channels, that is, channels that map quantum states to (finite-alphabet) classical probability distribution.
An important property of the measured R\'enyi divergences is that they are the minimal among quantum extensions of R\'enyi relative entropies that satisfy the data processing inequality: in particular, $D_\alpha^{\MM}(\rho\|\sigma) \leq \wt{D}_\alpha (\rho \| \sigma)$ for all $\alpha \in [1/2,\infty]$ and $D_\alpha^{\MM}(\rho\|\sigma) \leq \ol{D}_\alpha (\rho \| \sigma)$ for all $\alpha \in [0,2]$, with the inequalities generally strict~\cite{berta_2017-1}.

The variant of smooth relative entropies for quantum states that we focus on here are the \deff{measured smooth R\'enyi divergences}~\cite{regula_2026-1,regula_2026}, which are defined by optimising the corresponding classical smooth divergence over measurement channels:
\begin{equation}\begin{aligned}
 D_\alpha^{\ve,\MM} (\rho \| \sigma) \coloneqq \sup_{\M \in \MM} D^\ve_{\alpha}\! \left( \M(\rho) \middle\| \M(\sigma) \right).
\end{aligned}\end{equation}
The case of particular interest to us will be the measured collision divergence $D_2^{\MM}$, whose smooth form has recently been characterised by~\cite{regula_2026} and shown to admit variational expressions as
\begin{align}
  D_2^{\ve,\MM} (\rho \| \sigma) &= \log \inf \lset \norm{R}{\sigma}^2 \bar R = R^\dagger,\; \norm{\rho - R}{+} \leq \ve,\; R \leq \rho \rset \label{eq:D2_primal}\\
  &= \log \sup_{0 \leq W \leq \id}  \frac{\left(\Tr (W \rho) - \epsilon\right)_+^2}{\Tr W^2 \sigma}.
\end{align}
Unlike classical smooth relative entropies, the expression in~\eqref{eq:D2_primal} involves an optimisation over self-adjoint operators $R$ that need not be positive. 
Here, the \deff{Bures seminorm} is defined as
\begin{equation}\begin{aligned}
  \norm{A}{B} \coloneqq \begin{cases} \sqrt{\<A, A \>_B} & \; \text{if } A \in \ran(\J_{B}) \\ \infty & \; \text{otherwise,} \end{cases}
\end{aligned}\end{equation}
where the Bures inner product is 
\begin{equation}\begin{aligned}
  \< A, C \>_B \coloneqq \Tr A \J_B^{-1} (C), \qquad \J_{B} (X) \coloneqq \frac12 ( XB + BX ).
\end{aligned}\end{equation}
We see that $\J_{B} (X)$ can be understood as a matrix multiplication superoperator, and $\J_B^{-1} (X)$ thus plays the role of matrix division. The metric defined by the Bures inner product is the minimal operator-monotone Riemannian metric~\cite{petz_1996,lesniewski_1999,hiai_2012}.

From any R\'enyi divergence $\DD_\alpha \in \{ \ol{D}_\alpha, \wt{D}_\alpha, D_\alpha^{\MM}, D_\alpha^{\ve,\MM} \}$ and any CQ state $\rho_{XE}$, we define the optimised and unoptimised variants of conditional entropies as
\begin{equation}\begin{aligned}
  \HH_\alpha^{\uparrow}(X|E)_\rho &\coloneqq - \!\!\inf_{\substack{\sigma_E \geq 0\\\Tr \sigma_E = 1}} \DD_\alpha(\rho_{XE} \| \id_X \otimes \sigma_E),\\
  \HH_\alpha^{\downarrow}(X|E)_\rho &\coloneqq -\, \DD_\alpha(\rho_{XE} \| \id_X \otimes \rho_E).
\end{aligned}\end{equation}
When the side information system $E$ is trivial, this reduces to the smooth R\'enyi entropy $H_{\alpha}^\ve(X)_p = - D_\alpha^\ve(p_X \| \id_X )$, with $H_\alpha(X)_p = H_\alpha^0(X)_p$. 
We also follow the standard convention in using $H_{\min}$ to denote the conditional entropies constructed from the max-relative entropy, given by
\begin{equation}\begin{aligned}
  D_{\max}(\rho \| \sigma) \coloneqq \wt{D}_\infty(\rho\|\sigma) = \inf \lset \lambda \bar \rho \leq \exp(\lambda)\, \sigma \rset.
\end{aligned}\end{equation}


\subsection{Randomness extraction}

The setting of privacy amplification studied here is based on the notion of strong, seeded randomness extraction against quantum side information (see e.g.~\cite{tomamichel_2011}). It is implemented by a hash function $f \colon \X \to \Z$, whose random selection from  among a given family of functions is facilitated by an auxiliary system called a seed.

A family of hash functions $f \colon \X \to \Z$ is called \deff{$\boldsymbol{k}$-universal} if, for all distinct $x_1, \ldots, x_j \in \X$ with $2 \leq j \leq k$,
\begin{equation}\begin{aligned}
  \Pr_f \Big( f(x_1) = \ldots = f(x_j) \Big) \leq \frac{1}{|\Z|^{j-1}}.
\end{aligned}\end{equation}
If the above holds with equality, we call the family \deff{$\boldsymbol{k\*}$-universal}. Furthermore,  when $|\X|\geq k$, a family of hash functions is called \deff{strongly $\boldsymbol{k}$-universal} (or \deff{$\boldsymbol{k}$-independent}) if, for all distinct $x_1, \ldots, x_k \in \X$ and all $z_1, \ldots, z_k \in \Z$, it holds that
\begin{equation}\begin{aligned}
  \Pr_f \left( \bigwedge_{i=1}^k f(x_i) = z_i \right) = \frac{1}{|\Z|^k}.
\end{aligned}\end{equation}
Equivalently, strong $k$-universality means that $(f(x))_x$ are uniformly distributed, $k$-wise independent random variables. 
Every strongly $k$-universal family of hash functions is $k\*$-universal, but the converse need not be true --- for example, the commonly encountered hash family based on linear Toeplitz hashing is $2\*$-universal, but not strongly 2-universal~\cite{carter_1979,krawczyk_1994}. 

We will use the notation $\univ{k}$ and $\univstar{k}$ to refer to the sets of all $k$-universal and $k$\*-universal families of hash functions, respectively.

The performance of privacy amplification is measured in terms of the expected distance to perfect uniform randomness, independent of the adversary's system $E$. Here we quantify this with the trace distance, and so we say that privacy amplification is realised with error $\ve$ if
\begin{equation}\begin{aligned}
  \EE_f \frac12 \norm{\rho_{ZE}^f - \frac{\id_Z}{|\Z|} \otimes \rho_E}{1} \leq \ve,
\end{aligned}\end{equation}
where
\begin{equation}\begin{aligned}
  \rho^f_{ZE} \coloneqq \sum_z \proj{z} \otimes \!\!\sum_{x : f(x) = z} \rho_{E,x}
\end{aligned}\end{equation}
for a given initial input state $\rho_{XE} = \sum_x \proj{x} \otimes \rho_{E,x}$. 
Alternatively, explicitly accounting for the seed system $S$, the error can be understood as measuring the distance $\frac12 \norm{ \rho_{ZES} - \frac{\id_Z}{|\Z|} \otimes \rho_E \otimes \rho_S }{1}$. 
Our aim then becomes to minimise this distance and to understand its behaviour under universal hash functions.


\subsection{On converse bounds and optimality}

A question that motivates many of the results in this work is: how to argue the optimality of a bound on randomness extraction?

In the literature on cryptographic randomness extractors, one often encounters converse bounds that apply to all possible sources with a given min-entropy (see e.g.~\cite{vadhan_2012,berta_2017-2}). Here we are interested in a more  precise understanding of the performance of randomness extractors for a particular source. We will thus adopt an approach inspired by information-theoretic bounds that aim to measure the optimal extractable randomness of a given source state $\rho_{XE}$. 

One may then wish to establish a converse bound that applies to \emph{all} possible hash functions, in the sense of obtaining a lower bound on the \deff{best-case error}
\begin{equation}\begin{aligned}\label{eq:def_converse_allhash}
  \Delta_{\min} (\rho_{XE},\, |\Z|) \coloneqq \min_{f \colon \X \to \Z} \,\frac12 \norm{\rho_{ZE}^f - \frac{\id_Z}{|\Z|} \otimes \rho_E}{1}.
\end{aligned}\end{equation}
For a fixed source $\rho_{XE}$, this is the strongest possible converse bound: it tells us that even if we know the state exactly, we can do no better, no matter what hash function we choose. However, already in the classical case of randomness extraction with no side information, Hayashi~\cite{hayashi_2013} observed that such converse bounds may be too restrictive when applied to the error measured with trace distance, and the asymptotic behaviour of the quantity $\Delta_{\min}$ differs from the error achievable with random hash functions. (We will discuss this in detail in Section~\ref{sec:warmup}.) The issue is even more pronounced in the case of quantum side information, where the converse bounds that apply to all hash functions can be strictly weaker than classical ones~\cite{renes_2018,regula_2026}. 

Since in practice --- and especially when dealing with side information --- randomness extraction is not performed with source-dependent hashes but rather with universal ones, we instead aim to more appropriately characterise the best achievable error in such a context.
There is, however, still an ambiguity in what a converse bound is supposed to tell us. As achievability statements such as the leftover hash lemma are intended to work for \emph{any} family of hash functions satisfying some universality assumptions, one way to verify their optimality is to simply produce a single family which achieves the bound. This may be thought of as lower bounding a worst-case average error
\begin{equation}\begin{aligned}\label{eq:def_converse_max}
  \Delta_{\max,\frF} (\rho_{XE},\, |\Z|) \coloneqq \sup_{\F \in \frF}\, \EE_{f \in \F}\, \frac12 \norm{ \rho^f_{ZE} - \frac{\id_Z}{\left|\Z\right|} \otimes \rho_E }{1},
\end{aligned}\end{equation}
where $\F$ denotes the family of functions that our hash is sampled from, and $\frF$ refers to a chosen type of hash functions, such as $k$-universal ones. 
Converse bounds of this type were employed in particular by Hayashi and Watanabe~\cite{hayashi_2013,hayashi_2016-3} to argue the optimality of certain randomness extraction bounds. 
A potential concern is that such a bound may again not accurately reflect the performance that would be achieved in practice: for example, the family $\F$ of hash functions employed in~\cite{hayashi_2013,hayashi_2016-3} is based on randomly sampling a function from among all possible functions $\X \to \Z$, which is prohibitively difficult to realise in practice and thus represents more of an idealised limit. One may then instead bound the \deff{restricted best-case error}, namely the best achievable average error within a given family of hashes:
\begin{equation}\begin{aligned}\label{eq:def_converse_min}
  \Delta_{\min,\frF} (\rho_{XE},\, |\Z|) \coloneqq \inf_{\F \in \frF} \,\EE_{f \in \F}\, \frac12 \norm{ \rho^f_{ZE} - \frac{\id_Z}{\left|\Z\right|} \otimes \rho_E }{1}.
\end{aligned}\end{equation}
This resembles the idea of `ensemble converses'~\cite{gallager_1973} in information theory, where the performance of channel coding is evaluated under some restrictions on the allowed codes.

In the sequel we will give precise conditions on the applicability of all achievability and converse bounds we derive, and our results can be interpreted as either certifying the optimality in the sense of~\eqref{eq:def_converse_max} or as establishing ensemble converses in the sense of~\eqref{eq:def_converse_min}.

A particularly important sense in which we wish to understand the optimality of achievability results for randomness extraction is as bounds on the \emph{scaling} of the error.
What we mean by this precisely is the scaling in the number of copies when applied to many independent and identically distributed (i.i.d.) copies of the given source state, $\rho_{XE}^{\otimes n}$.
When extracting randomness of size $|\Z| = \exp(nR)$ at a rate $R < H(X|E)_\rho$, the decay of the error in such an i.i.d.\ scenario is generally exponential in $n$; as such, showing the optimality of a bound up to a constant is perfectly sufficient for the purposes of determining its tightness. 
While the assumption of an i.i.d.\ source is a somewhat idealised one, in practice the analysis of randomness extraction and its applications is typically reduced precisely to i.i.d.\ or closely related cases (as often seen e.g.\ in quantum key distribution~\cite{renner_2005,christandl_2009,dupuis_2020}), and the i.i.d.\ analysis thus provides meaningful information about the many-copy behaviour of practical bounds.
In information-theoretic terms, the optimal rate of decay of the error of extracting randomness from $n$ i.i.d.\ copies of a source, understood in an asymptotic sense as $\lim_{n\to\infty} -\frac1n \log\!\left[\Delta_{\min,\frF}(\rho^{\otimes n}_{XE}, \exp(nR))\right]$, is known as the error exponent (reliability function); if, on the other hand, the demanded randomness rate is too large, namely $R > H(X|E)_\rho$, then the error will converge to 1, and the corresponding rate of convergence is known as the strong converse exponent.


\section[Warm-up: Intrinsic randomness with no side information]{Warm-up:\\*Intrinsic randomness with no side information}
\label{sec:warmup}

To build some intuition and see how connections with hypothesis-testing--style quantities naturally emerge in the analysis of randomness extraction, we consider first the classical case of randomness extraction without side information; this is often studied as the problem of intrinsic randomness~\cite{vembu_1995-1}, where the extraction is performed with source-dependent hash functions. Through a careful estimation of the involved problems, we will easily recover tight asymptotic bounds and develop one-shot relations that improve on some prior results.

Readers interested in the case of quantum side information may skip directly to Section~\ref{sec:universal}.

\subsection{Information spectrum and hypothesis testing}

In the study of intrinsic randomness, we are interested in finding a function $f\colon \X \to \Z$ which minimises
\begin{equation}\begin{aligned}
 \Delta_{\min}(p_X, |\Z|) = \inf_f \norm{ p^f_Z - \frac{\id}{|\Z|}}{+},
\end{aligned}\end{equation}
where $p^f_Z(z) = \sum_{x: f(x) = z} p_X(x)$ and we recall that $\| \cdot \|_+$ denotes the generalised trace distance. Clearly, what the function $f$ is trying to do is to spread the elements of $p_X$ into bins equally, so that each bin has mass at most $1/|\Z|$; since this is not always possible depending on the given $p_X$, we need to bound how much we exceed the desired bin size. 

A very easy bound is: we can always distribute the elements of $p_X$ into $|\Z|$ bins where we stop filling a bin as soon as we exceed $1/|\Z|$. This means that each bin exceeds $1/|\Z|$ by at most one extra element, giving the estimate $p^f_Z(z) \leq \frac{1}{|\Z|} + \max_x p_X(x)$. This is not very good in general, as the largest element can dominate the bound. The way to mitigate this is exactly smoothing:  divide $p_X = q_X + r_X$, where $q$ and $r$ are unnormalised distributions, and construct the function $f$ by applying the procedure to $r_X$, rather than $p_X$. The triangle inequality then gives
\begin{equation}\begin{aligned}
  \norm{ p^f_Z - \frac{\id}{|\Z|}}{+} &\leq  \norm{p^f_Z - r^f_Z}{+} + \Tr \!\left( r^f_Z - \frac{\id}{|\Z|}\right)_+ \\
  &\leq   \norm{p_X - r_X}{+} + |\Z| \, \max_x r_X(x).
\end{aligned}\end{equation}
Recalling that
\begin{equation}\begin{aligned}
  H^\ve_{\min}(X)_p = - \log \inf \lset \max_x r_X(x) \bar 0 \leq r_X \leq p_X,\; \norm{p_X - r_X}{+} \leq \ve \rset,
\end{aligned}\end{equation}
we can rewrite the above to get a simple but fundamental bound:
\begin{boxed}[white]
\begin{lemma}[Non-universal min-entropy bound]\label{lem:minent}
For all $\ve \in [0,1)$, \vspace*{-.2\baselineskip}
\begin{equation}\begin{aligned}
  \Delta_{\min}(p_X, |\Z|) \leq \ve + |\Z| \exp\!\left(- H^\ve_{\min}(X)_p\right).
\end{aligned}\end{equation}
\end{lemma}
\end{boxed}
Another way to understand this is to use the fact that $H_{\min}(X)_r = - \log \inf \lset \lambda \bar r \leq \lambda \id \rset$, which, upon a minimisation over $\ve$, lets us rewrite
\begin{equation}\begin{aligned}
 \inf_{\ve} \Big[ \ve + |\Z| \exp\!\left(- H^\ve_{\min}(X)_p\right) \Big] &= \inf_{\lambda\geq0} \Big[ \Tr (p_X - \lambda)_+ + |\Z| \, \lambda \Big]\\
 &\leq \inf_{\lambda\geq0} \Big[ \Tr p_X \big\{ p_X \geq \lambda \} + |\Z| \,\lambda \Big].
\end{aligned}\end{equation}
Here in the second line we obtain the information spectrum bound that Han~\cite{han_2003} and later Hayashi~\cite{hayashi_2013,watanabe_2013} used as a one-shot bound on intrinsic randomness; we see that the min-entropy bound is actually tighter.

Sticking with the min-entropy bound, we can rewrite the optimisation in a convenient way: Lagrange duality gives that
\begin{equation}\begin{aligned}
  \inf_{\ve} \Big[ \ve + |\Z| \exp\!\left(- H^\ve_{\min}(X)_p\right) \Big] &= \inf \lset \Tr(p_X - r_X) + |\Z|\, \lambda \bar 0 \leq r_X \leq \lambda \id_X,\; r_X \leq p_X \rset\\
  &= \sup \lset \Tr M_X p_X \bar 0 \leq M_X \leq \id,\; \Tr M_X \leq |\Z| \rset.
\end{aligned}\end{equation}
One can see that this in fact corresponds to a hypothesis testing problem. Precisely, when the type I error is constrained to be at most $\ve$, let us define 
the (type II) success probability in hypothesis testing between two positive semidefinite operators $B$ and $A$ as
\begin{equation}\begin{aligned}
  \pi_\ve(A \| B) &\coloneqq \Tr A - \beta_\ve(B\| A) \\
  &\hphantom{:}= \sup \lset \Tr M A \bar 0 \leq M \leq \id,\; \Tr M B \leq \ve \rset.
\end{aligned}\end{equation}
We note here that, for notational convenience, our definition of $\pi_\ve$ interchanges the roles of $A$ and $B$ compared to standard notation for the type II error $\beta_\ve$. 
Using this expression, we then have the following.

\begin{boxed}[white]
\begin{corollary}[Hypothesis testing bound]\label{cor:hyp_test_bound}
The min-entropy bound of Lemma~\ref{lem:minent} is equivalent to the bound\vspace*{-.3\baselineskip}
\begin{equation}\begin{aligned}\label{eq:hyp_test_bound}
  \Delta_{\min}(p_X, |\Z|) \leq \pi_{|\Z|}(p_X \| \id_X) = \pi_{|\Z|/|\X|}\left( p_X \middle\| \frac{\id_X}{|\X|}\right).
\end{aligned}\end{equation}
\end{corollary}
\end{boxed}

From this, one can immediately derive various other bounds, both one-shot and asymptotic. Of interest is for example the one-shot R\'enyi bound
\begin{equation}\begin{aligned}
  \pi_{|\Z|}\left(  p_X \middle\| \id_X \right) &\leq |\Z|^{\frac{\alpha-1}{\alpha}} \exp\!\left(- \frac{\alpha-1}{\alpha} H_\alpha(X)_p\right)
\end{aligned}\end{equation}
for all $\alpha > 1$, which is a simple consequence of the data-processing inequality for $D_\alpha$. However, the i.i.d.\ exponent bound we get from this, with $|\Z^n| = \exp(n R)$, is
\begin{equation}\begin{aligned}
  \liminf_{n\to\infty} - \frac1n \log \left[ \Delta_{\min}(p^{\otimes n}_{X^n}, \exp(nR)) \right] \geq \sup_{\alpha > 1} \frac{\alpha-1}{\alpha} \left( H_{\alpha}(X)_p - R \right).
\end{aligned}\end{equation}
This is tight as a bound on the asymptotics of hypothesis testing, but it does not optimally characterise the exponent of randomness extraction.

\subsection{Collision without leftover hashing}

To look for an improved bound, clearly we need a tighter way of counting the mass of the elements of $p^f_Z$ that exceed  $\frac{1}{|\Z|}$. To this end, we will consider a simple randomised binning approach as follows. 

Start with a (subnormalised) distribution $r_X$. Fill the first bin with elements $\{1, \ldots, x_1 \} \in \X$ until $\sum_{x' \leq x_1} r_X(x') > \frac{1}{|\Z|}$; the second bin with elements $\{ x_1 + 1, \ldots, x_2 \}$ until $\sum_{x' \leq x_2} r_X(x') > 2\frac{1}{|\Z|}$, et cetera, in the sense that each $x$ is assigned to the bin $z$ for the least $z \in \{ 1, \ldots, |\Z| \}$ such that $\sum_{x' < x} r_X(x') \leq z \frac{1}{|\Z|}$. 
An element $x$ thus exceeds the $z$th bin if $f(x)=z$ and $\sum_{x'<x} r_X(x') \leq z\tfrac{1}{|\Z|} < \sum_{x' \leq x} r_X(x')$. 
Counting up the total mass of such elements gives the estimate
\begin{equation}\begin{aligned}
  \norm{ p^f_Z - \frac{\id}{|\Z|}}{+} &\leq  \norm{p^f_Z - r^f_Z}{+} + \sum_z \left( \sum_{x: f(x) = z} r_X(x) - \frac{1}{|\Z|} \right)_+\\
  &\leq \norm{p_X - r_X}{+} +  \sum_z \sum_{x} r_X(x) \, \boldsymbol{1}\! \left\{ x \text{ exceeds the $z$th bin }\! \right\}\\
  &=  \norm{p_X - r_X}{+} +  \sum_{x} r_X(x) \, \boldsymbol{1}\! \left\{ x \text{ exceeds a bin }\! \right\}.
\end{aligned}\end{equation}
A way to bound this quantity is to randomise the selection of bins. Consider a modified binning function $f_t : \X \to \Z$ that introduces an initial offset $t$: 
each element $x'$ is assigned to the bin $z$ for the least $z$ such that $\sum_{x \leq x'} r_X(x) \leq z \frac{1}{|\Z|} + t$ (modulo 1). 
Taking a uniform average over all such starting offsets $t \in \big[0, \tfrac{1}{|\Z|}\!\big]$, we can count the fraction of bins that a given $x$ exceeds as
\begin{equation}\begin{aligned}
  \Pr_{t} \left( x \text{ exceeds a bin } \right) &= \Pr_{t} \Big( r_X(x) \notin [0, t] \Big)\\
  &= \frac{1}{1/|\Z|} \int_{0}^{1/|\Z|} \boldsymbol{1}\!\left\{ r_X(x) > t \right\} \mathrm{d}t\\
  &= \min \left\{ 1,\, |\Z|\, r_X(x) \right\}.
\end{aligned}\end{equation}
Thus
\begin{equation}\begin{aligned}
  \EE_{t} \norm{ p^{f_t}_Z - \frac{\id}{|\Z|}}{+} &\leq  \norm{p_X - r_X}{+} + \EE_t \sum_{x} r_X(x) \, \boldsymbol{1}\! \left\{ x \text{ exceeds a bin }\! \right\}\\
  &\leq \norm{p_X - r_X}{+} + \sum_{x} r_X(x) \, \Pr_t \!\Big( x \text{ exceeds a bin }\! \Big)\\
  &\leq \norm{p_X - r_X}{+} + |\Z| \sum_{x} r_X(x)^2.
\end{aligned}\end{equation}
This implies in particular that there exists a choice of hash function $f_t$ which achieves $ \norm{ p^{f_t}_Z - \frac{\id}{|\Z|}}{+} \leq \norm{p_X - r_X}{+} + |\Z| \sum_{x} r_X(x)^2$, giving the following bound.

\begin{tboxed}[white]{Collision-style achievability bound for intrinsic randomness}
\begin{proposition}\label{prop:nonuniv_collision_bound}
For all $\ve \in [0,1)$, 
\begin{equation}\begin{aligned}\label{eq:nonuniv_collision_bound}
  \Delta_{\min}(p_X, |\Z|) &\leq \ve + |\Z| \exp\!\left(- H^\ve_{2}(X)_p\right) \\
  &\leq \ve + |\Z| \, (1-\ve) \exp\!\left(- H^\ve_{\min}(X)_p\right).
\end{aligned}\end{equation}
Equivalently, for any $\lambda$ it holds that
\begin{equation}\begin{aligned}\label{eq:secondline}
  \Delta_{\min}(p_X, |\Z|) &\leq \Tr(p_X - \lambda)_+ + |\Z| \Tr (p_X \wedge \lambda)^2.
\end{aligned}\end{equation}
\end{proposition}
\end{tboxed}
Note that the standard result of the leftover hash lemma~\cite{impagliazzo_1989,bennett_1995} would instead give the bound $\Delta_{\min}(p_X) \leq \ve + \frac12 \sqrt{|\Z| \exp\!\left(- H^\ve_{2}(X)_p\right)}$. This in general exhibits a different scaling and would lead to a different exponent; see Section~\ref{sec:universal}.

The rewriting in~\eqref{eq:secondline} follows since $D^\ve_2(p\|q) = D_2(p'\|q)$ where $p' = p \wedge \lambda q$ with $\Tr (p - \lambda q)_+ = \ve$~\cite{yang_2019}, and in this case $q = \id$.

Taking the Lagrange dual of the optimisation in~\eqref{eq:nonuniv_collision_bound}, we have that
\begin{equation}\begin{aligned}\label{eq:collisionlike_dual}
  \inf_{0 \leq r_X \leq p_X} \left[ \norm{p_X - r_X}{+} + |\Z| \sum_{x} r_X(x)^2 \right] = \sup \lset \Tr M_X p_X - \frac{1}{4 |\Z|} \Tr M_X^2 \bar 0 \leq M_X \leq \id \rset,
\end{aligned}\end{equation}
and rearranging this shows the connection with `half-quadratic' hypothesis testing: precisely,
\begin{equation}\begin{aligned}
  1 - \sup_{0 \leq M_X \leq \id} \left[ \Tr M_X p_X - \frac{1}{4 |\Z|} \Tr M_X^2  \right] = \inf_{0 \leq M_X \leq \id} \left[ \Tr (\id - M_X) p_X + \frac14 \frac{|\X|}{|\Z|} \Tr M^2_X \frac{\id_X}{|\X|} \right],
\end{aligned}\end{equation}
which is akin to optimising the average error in hypothesis testing, but with one of the measurement operators squared. This will indeed turn out to be a recurring theme in all settings studied in this paper, and we will return to such problems shortly.

In fact, we can use the dual form of~\eqref{eq:collisionlike_dual} to easily get an analytical form: on the one hand, $r_X =  p_X \wedge \frac{\id_X}{2 |\Z|}$ is feasible for the primal, and on the other hand $M_X = \id_X \wedge 2 |\Z| p_X$ is feasible for the dual, giving
\begin{equation}\begin{aligned}
  \inf_\ve \Big[ \ve + |\Z| \exp\!\left(- H^\ve_{2}(X)_p\right)\big] = \sum_{x: p_X(x) \geq \frac{1}{2 |\Z|}} \left( p_X(x) - \frac{1}{4 |\Z|} \right) +  |\Z|  \sum_{x: p_X(x) < \frac{1}{2 |\Z|}} p_X(x)^2.
\end{aligned}\end{equation}
To make more intuitive sense of this, it is actually easier to pick suboptimal solutions: $r_X =  p_X \wedge \frac{\id_X}{|\Z|}$ for the primal, and $M_X = \id_X \wedge |\Z| p_X$ for the dual. Plugging these solutions into the optimisation and defining $S = \left\{ x : p_X(x) \geq  \frac{1}{|\Z|} \right\}$ and $S^\perp = \left\{ x : p_X(x) < \frac{1}{|\Z|} \right\}$, we have that
\begin{align}
  \sum_{x \in S} p_X(x) +  |\Z|  \sum_{x \in S^\perp} p_X(x)^2 &= \norm{p_X - r_X}{+} + |\Z| \sum_{x} r_X(x)^2 \nonumber\\
  &\geq \inf_\ve \Big[ \ve + |\Z| \exp\!\left(- H^\ve_{2}(X)_p\right)\big] \nonumber\\
  &\geq \Tr M_X p_X - \frac{1}{4 |\Z|} \Tr M_X^2\\
  &= \sum_{x \in S} p_X(x) + |\Z| \sum_{x \in S^\perp} p_X(x)^2 - \frac{1}{4|\Z|} \left( \sum_{x\in S} 1 + \sum_{x\in S^\perp} |\Z|^2 \,p_X(x)^2 \right) \nonumber\\
  &\geq \sum_{x \in S} p_X(x) + |\Z| \sum_{x \in S^\perp} p_X(x)^2  - \frac{1}{4 |\Z|}  \left( \sum_{x\in S} |\Z| \,p_X(x) + \sum_{x\in S^\perp} |\Z|^2\, p_X(x)^2 \right) \nonumber\\
  &= \frac34 \left(  \sum_{x \in S} p_X(x) + |\Z| \sum_{x \in S^\perp} p_X(x)^2 \right), \nonumber
\end{align}
which means that we are losing very little with these suboptimal choices. This gives a more convenient bound:
\begin{boxed}[white]
\begin{corollary}[Slightly relaxed collision-style bound]
\begin{equation}\begin{aligned}\label{eq:slightly_relaxed}
  \Delta_{\min}(p_X, |\Z|) \leq  \sum_{x \in S} p_X(x) +  |\Z|  \sum_{x \in S^\perp} p_X(x)^2  = 
  \Tr \!\left( p_X \wedge |\Z| p_X^2 \right),
\end{aligned}\vspace*{-.5\baselineskip}\end{equation}
with this bound being equivalent up to a constant factor to the bound of Proposition~\ref{prop:nonuniv_collision_bound}.
\end{corollary}
\end{boxed}
From this, we immediately get an interesting consequence: a one-shot bound in terms of R\'enyi divergences that is asymptotically tight.

\begin{boxed}[white]
\begin{corollary}[Exponential bound on intrinsic randomness]\label{cor:collision_exp_noside}
For all $\alpha \in (1,2]$, it holds that
\begin{equation}\begin{aligned}
  \Delta_{\min}(p_X, |\Z|) \leq |\Z|^{\alpha-1} \exp\!\big(-(\alpha-1) \,H_\alpha(X)_p\big).
\end{aligned}\end{equation}
Consequently, 
\begin{equation}\begin{aligned}
  \liminf_{n\to\infty} - \frac1n \log \left[ \Delta_{\min}(p^{\otimes n}_{X^n}, \exp(nR)) \right] \geq \sup_{\alpha \in (1,2]} \,(\alpha-1) \Big( H_{\alpha}(X)_p - R \Big),
\end{aligned}\end{equation}
which recovers the asymptotic result of Hayashi~\cite[Lemma~5]{hayashi_2013}.
\end{corollary}
\end{boxed}
\begin{proof}
Let $S = \left\{ x : p_X(x) \geq  \frac{1}{|\Z|} \right\}$ and $S^\perp = \left\{ x : p_X(x) < \frac{1}{|\Z|} \right\}$ as before. 
The bound of Eq.~\eqref{eq:slightly_relaxed} gives
\begin{align}
  \Delta_{\min}(p_X) & \leq \sum_{x \in S} p_X(x) + \sum_{x \in S^\perp} |\Z|\, p_X(x)^2 \nonumber\\
  &= \sum_{x \in S} p_X(x)^{\alpha} \,p_X(x)^{-(\alpha-1)}  + \sum_{x \in S^\perp} |\Z| \,p_X(x)^\alpha p_X(x)^{2-\alpha}\\
  &\leq \sum_{x \in S} p_X(x)^{\alpha} \,|\Z|^{\alpha-1}  + \sum_{x \in S^\perp} |\Z|\, p_X(x)^\alpha \,|\Z|^{\alpha-2}\nonumber\\
  &= |\Z|^{\alpha-1} \sum_x p_X(x)^\alpha. \tag*{\qedhere}
\end{align}
\end{proof}

\subsection{Error exponent and strong converse exponent}

A standard min-entropy converse bound (see e.g.~\cite{renner_2005-2,watanabe_2013,yang_2019}) tells us that
\begin{equation}\begin{aligned}
    \Delta_{\min}(p_X, |\Z|) \geq \Tr\!\left( p_X - \frac{\id_X}{|\Z|} \right)_+.
\end{aligned}\end{equation}
From this, an application of standard large-deviation bounds such as Cram\'er's or Sanov's theorem tells us that, with $|\Z^n| = \exp(n R)$, we have that
\begin{equation}\begin{aligned}\label{eq:converse_asymp_noside}
  \limsup_{n\to\infty} - \frac1n \log \left[ \Delta_{\min}(p^{\otimes n}_{X^n}, \exp(nR)) \right] \leq \sup_{\alpha > 1} \,(\alpha-1) \Big( H_{\alpha}(X)_p - R \Big)
\end{aligned}\end{equation}
for all $R>H_{\min}(X)_p$ (with the exponent diverging to infinity for smaller rates). 
Together with the achievability result in Corollary~\ref{cor:collision_exp_noside}, this gives a tight exponent of intrinsic randomness (non-universal randomness extraction) for high enough rates, matching the observation of~\cite{hayashi_2013}. 
\begin{boxed}[white]
\begin{corollary}[{Error exponent of intrinsic randomness~\cite[Theorem~4]{hayashi_2013}}]
For any $R > R_{\rm crit} \coloneqq \left.\frac{\mathrm{d}}{\mathrm{d}\alpha} (\alpha-1) H_\alpha(X)_p \right|_{\alpha=2}$, the supremum in~\eqref{eq:converse_asymp_noside} is achieved for $\alpha \in (1,2]$, and hence
\begin{equation}\begin{aligned}
   \lim_{n\to\infty} - \frac1n \log \left[ \Delta_{\min}(p^{\otimes n}_{X^n}, \exp(nR)) \right] =  \sup_{\alpha \in (1,2]} \,(\alpha-1) \Big( H_{\alpha}(X)_p - R \Big).
\end{aligned}\end{equation}
\end{corollary}
\end{boxed}

From the collision-style one-shot bound in Proposition~\ref{prop:nonuniv_collision_bound}, one can also get an achievability for the strong converse exponent. This bound is in fact tight, as recently shown in~\cite{berta_2025-1}. 
\begin{boxed}[white]
\begin{corollary}[{Strong converse exponent of intrinsic randomness~\cite[Theorem~17]{berta_2025-1}}]
\begin{equation}\begin{aligned}
  \lim_{n\to\infty} - \frac1n \log \left[ 1 - \Delta_{\min}(p^{\otimes n}_{X^n}, \exp(nR)) \right] = \sup_{\alpha \in (0,1)} \,(\alpha-1) \Big( H_{\alpha}(X)_p - R \Big).
\end{aligned}\end{equation}
\end{corollary}
\end{boxed}
\begin{remark}
The same exponent can in fact be achieved by 2-universal hash functions. We will return to this later in Theorem~\ref{thm:strongconverse}.
\end{remark}
\begin{proof}
From Proposition~\ref{prop:nonuniv_collision_bound} we have that $ \Delta_{\min}(p_X, |\Z|) \leq \ve + |\Z| \, (1-\ve) \exp\!\left(- H^\ve_{\min}(X)_p\right)$, and hence
\begin{equation}\begin{aligned}
   1 - \Delta_{\min}(p_X, |\Z|) &\geq 1 - \ve - |\Z| \, (1-\ve) \exp\!\left(- H^\ve_{\min}(X)_p\right).
\end{aligned}\end{equation}
Taking $\ve = 1 - \exp(-n T_n)$ and $|\Z| = \exp ( n R )$, this gives
\begin{equation}\begin{aligned}
  1 - \Delta_{\min}(p^{\otimes n}_{X^n}, \exp(nR)) \geq \exp(-n T_n) - \exp\left(- n T_n + n R - H^{1-\exp(-n T_n)}_{\min}(X^n)_{p^{\otimes n}}\right).
\end{aligned}\end{equation}
Using now the known strong converse exponent of $D^\ve_{\max}$ --- which follows from Cram\'er's theorem for large deviations, with an explicit expression e.g.\ in~\cite[Theorem~2]{salzmann_2022} ---
we have that there exists a choice of $T_n$ such that ${\displaystyle \lim_{n\to\infty}} \frac1n H^{1-\exp(-n T_n)}_{\min}(X^n)_{p^{\otimes n}} = R$ and $ {\displaystyle \lim_{n\to\infty} T_n} = \sup_{\alpha \in (0,1)} \,(\alpha-1) \Big( H_{\alpha}(X)_p - R \Big)$, 
which gives the claimed achievability result. 
The equality holds because this matches the converse bound of~\cite[Theorem~17]{berta_2025-1} or~\cite[Proposition 26]{regula_2026}.
\end{proof}

Most of the results above, and in particular the tightest one-shot bounds, make very explicit use of the knowledge of the underlying source distribution. Although this will provide useful intuition about the structure of randomness extraction, the need for source-dependent hash functions is not in line with how privacy amplification is performed in practical contexts such as cryptography, and it also makes many of the results difficult to generalise to the case when quantum side information is involved. 
We thus move on to the discussion of the case of general sources with quantum side information, from which randomness is extracted using universal hash functions.

\section{Privacy amplification with universal hash functions}
\label{sec:universal}

\subsection{A different look at the leftover hash lemma}
\label{sec:different_look_lhl}

Recall that we defined the success probability of standard hypothesis testing as
\begin{equation}\begin{aligned}
  \pi_\ve(A \| B) &= \sup \lset \Tr M A \bar 0 \leq M \leq \id,\; \Tr M B \leq \ve \rset.
\end{aligned}\end{equation}
Let us then define the probability of success of \deff{half-quadratic hypothesis testing} as
\begin{equation}\begin{aligned}
  \hqr\pi_\ve (A \| B) &\coloneqq \sup \lset \Tr M A \bar 0 \leq M \leq \id,\; \Tr M^2 B \leq \ve \rset.
\end{aligned}\end{equation}
Note that the quadratic constraint is only applied to the alternative hypothesis $B$. It will be convenient to observe that $\hqr\pi_\ve ( k A \| m B) = k\, \hqr\pi_{\ve/m} ( A \| B)$ for $k,m > 0$. 

We now show the smooth leftover hash lemma can be understood exactly in terms of such a half-quadratic hypothesis testing problem, giving somewhat of a 2-universal analogue of the one-shot hypothesis testing bound that we saw in Eq.~\eqref{eq:hyp_test_bound}.

\begin{tboxed}[filled]{Leftover hash lemma as a hypothesis testing problem}
\begin{lemma}\label{lem:lhl}
The leftover hash lemma of~\cite[Theorem~14]{regula_2026} is equivalent to the statement that, for any classical--quantum $\rho_{XE} \geq 0$ and any 2\*-universal family of hash functions $f \colon \X \to \Z$, it holds that
\begin{align}
  \EE_f \frac12 \norm{\rho_{ZE}^f - \frac{\id_Z}{|\Z|} \otimes \rho_E}{1} &\leq 
  \inf_{\sigma_E} \; \hqr\pi_{|\Z|/4}  \big(  \rho_{XE} \big\| \id_X \otimes \sigma_E \big) \label{eq:lhl_different_forms1}\\
  &= \sup \lset \Tr M_{XE} \rho_{XE} \bar 0 \leq M_{XE} \leq \id,\; \norm{\Tr_X M_{XE}^2}{\infty} \leq \frac{|\Z|}{4} \rset \label{eq:lhl_different_forms2}\\
  &= \inf_{R=R^\dagger} \left[ \norm{\rho_{XE} - R_{XE}}{1} + \frac12 \sqrt{|\Z|} \inf_{\sigma_E} \norm{R_{XE}}{\id_X \otimes \sigma_E} \right], \label{eq:lhl_different_forms3}
\end{align}
where we recall that the Bures seminorm $\norm{A}{B} = \sqrt{\<A, A \>_B}$ is defined using the inner product $\< A, C \>_B = \Tr A \J_B^{-1} (C)$ with $\J_{A} (B) = \frac12 ( AB + BA )$. 
Here, the infimum over $\sigma_E$ denotes an optimisation over all quantum states, i.e. $\sigma_E \geq 0$, $\Tr \sigma_E = 1$. 
The operators $M_{XE}$ and $R_{XE}$ involved in the optimisation can always be taken to be classical--quantum.

\smallskip Furthermore, a relaxed statement of this relation holds also if the family of hash functions is only 2-universal. The result of~\cite[Proposition~7]{tan_2026} then corresponds to
\begin{equation}\begin{aligned}\label{eq:lhl_2univ_different_forms}
  \EE_f \frac12 \norm{\rho_{ZE}^f - \frac{\id_Z}{|\Z|} \otimes \rho_E}{1} &\leq 
  \inf_{\sigma_E} \; \hqr\pi_{|\Z|/2}  \big(  \rho_{XE} \big\| \id_X \otimes \sigma_E \big) \\
  &= \sup \lset \Tr M_{XE} \rho_{XE} \bar 0 \leq M_{XE} \leq \id,\; \norm{\Tr_X M_{XE}^2}{\infty} \leq \frac{|\Z|}{2} \rset\\
  &= \inf_{R=R^\dagger} \left[ \norm{\rho_{XE} - R_{XE}}{1} + \frac{1}{\sqrt2} \sqrt{|\Z|} \inf_{\sigma_E} \norm{R_{XE}}{\id_X \otimes \sigma_E} \right].
\end{aligned}\end{equation}
\end{lemma}
\end{tboxed}
\begin{proof}
Theorem~14 of~\cite{regula_2026} tells us that, for any $\ve \in [0,1)$ and any $\sigma_E$, the expected distance under 2*-universal hash functions is bounded as
\begin{equation}\begin{aligned}
  \EE_f \frac12 \norm{\rho_{ZE}^f - \frac{\id_Z}{|\Z|} \otimes \rho_E}{1} \leq \ve + \frac12 \sqrt{ |\Z| \, Q_2^{\ve,\MM} \left( \rho_{XE} \middle\| \id_X \otimes \sigma_E \right) },
\end{aligned}\end{equation}
where $Q_2^{\ve,\MM}(\rho\|\sigma) = \exp( \Ds{2}{\ve}{M}(\rho\|\sigma) )$, whose variational form we recall as
\begin{equation}\begin{aligned}
  Q_2^{\ve,\MM} (\rho \| \sigma) &= \inf \lset \norm{R}{\sigma}^2 \bar R \leq \rho, \; \norm{\rho - R}{+} \leq \ve \rset.
\end{aligned}\end{equation}
This can be extended to 2-universal hash functions at the expense of an additional factor of 2 inside the square root~\cite[Proposition~7]{tan_2026}, giving
\begin{equation}\begin{aligned}
  \EE_f \frac12 \norm{\rho_{ZE}^f - \frac{\id_Z}{|\Z|} \otimes \rho_E}{1} \leq \ve + \frac{1}{\sqrt{2}} \sqrt{ |\Z| \, Q_2^{\ve,\MM} \left( \rho_{XE} \middle\| \id_X \otimes \sigma_E \right) }.
\end{aligned}\end{equation}
To get the best upper bound, we thus need to minimise these expressions over all $\ve$ and over all $\sigma_E$. 
For simplicity let us denote $\sigma \coloneqq \id_X \otimes \sigma_E$ and, depending on the considered type of hash functions, either $k \coloneqq \frac{1}{4} |\Z|$ or $k \coloneqq \frac{1}{2} |\Z|$. Consider now that
\begin{equation}\begin{aligned}\label{eq:convolution}
  \inf_{\ve \in [0,1]} \left[ \ve + \sqrt{ k \, Q_2^{\ve,\MM} \left( \rho_{XE} \middle\| \sigma_{XE} \right) } \right] &= \inf \lset \norm{\rho - R}{+} + \sqrt{k} \, \norm{R}{\sigma} \bar R \leq \rho \rset\\
  &= \inf \lset \Tr ( \rho - R ) +\sqrt{k} \, \norm{R}{\sigma} \bar R \leq \rho \rset.
\end{aligned}\end{equation}
Writing $R = \J_\sigma(Z)$ for some $Z$, the Lagrange dual function of this problem evaluated at a Lagrange multiplier $P_{XE} \geq 0$ is
\begin{equation}\begin{aligned}
 \inf_{R=R^\dagger} L(R;\,P)  &= \inf_{R=R^\dagger}\, \Tr (\rho - R) + \sqrt{k} \sqrt{ \< R, R\>_{\sigma} } - \Tr P (\rho - R) \\
 &=  \inf_{Z=Z^\dagger} \, \Tr \left[\J_{\sigma}(Z) (P - \id)\right] + \sqrt{k} \sqrt{ \Tr \J_{\sigma}(Z) Z  } + \Tr (\id - P)\rho.
\end{aligned}\end{equation}
Identifying this as a quadratic form and using that $\displaystyle \inf_{a \in \RR} \big( \< a, b \> + m \norm{a}{} \big) = \smash{\begin{cases} 0 & \text{ if } \norm{b}{} \leq m \\ - \infty & \text{ if } \norm{b}{} > m\end{cases}}$ gives the Lagrange dual as
\begin{equation}\begin{aligned}\label{eq:lagdual}
  \sup_{P \geq 0} \inf_{R=R^\dagger} L(R;\,P) &= \sup \lset \Tr (\id - P)\rho \bar \sqrt{ \Tr \J_{\sigma}(\id - P)(\id- P)}  \leq  \sqrt{k} ,\; P \geq 0 \rset\\
  &= \sup \lset \Tr (\id - P)\rho \bar \Tr (\id - P)^2 \sigma \leq k,\; P \geq 0 \rset.
\end{aligned}\end{equation}
Notice now that, since $P \geq 0$, we have $(\id - P)_+ \in [0,\id]$, and so for any feasible $P$ it holds that $P' \coloneqq \id - (\id - P)_+ \in [0,\id]$. This operator is also feasible, since $(\id - P')^2 = (\id - P)_+^2 \leq (\id - P)^2_+ + (\id - P)^2_- = (\id - P)^2$. 
But any such feasible $P'$ can only improve the feasible optimal value: $(\id - P)_+ \geq (\id - P)$ implies that $P' \leq P$, and hence $\Tr (\id - P')\rho \geq \Tr (\id - P) \rho$.  Thus it suffices to optimize over $P \in [0,\id]$. Defining $M \coloneqq \id - P$ altogether gives
\begin{equation}\begin{aligned}\label{eq:variational}
 \inf_{\ve \in [0,1]} \left[ \ve +  \sqrt{ k \, Q_2^{\ve,\MM} \left( \rho_{XE} \middle\| \sigma_{XE} \right) } \right] &= \sup \lset \Tr M \rho \bar 0 \leq M \leq \id, \; \Tr M^2 \sigma \leq k \rset\\
 &= \hqr\pi_{k} (\rho\|\sigma).
\end{aligned}\end{equation}
Recalling that $\sigma=\id_X\otimes\sigma_E$, the quadratic constraint becomes $\Tr M^2 \sigma = \Tr \left[ (\Tr_X M_{XE}^2) \sigma_E \right]$. Since an optimisation over all states $\sigma_E$ gives $\sup_{\sigma_E} \!\Tr A_E \sigma_E = \|A_E\|_\infty$ for any $A_E \geq 0$, the variational form in Eq.~\eqref{eq:lhl_different_forms2} (or the corresponding expression in~\eqref{eq:lhl_2univ_different_forms}) follows. The equivalence with the hypothesis testing problem in~\eqref{eq:lhl_different_forms1} follows from a minimax argument applied to the quadratic constraint at each fixed lower bound on $\Tr(M_{XE}\rho_{XE})$, which proves that there exists one test $M_{XE}$ that satisfies the constraint for every $\sigma_E$.

The last expression in Eq.~\eqref{eq:lhl_different_forms3} simply removes the constraint $R \leq \rho$. The optimisation can then be rephrased as
\begin{equation}\begin{aligned}
  \inf_R \left[ \norm{\rho  - R}{1} + \sqrt{k} \norm{R_{XE}}{\sigma} \right] &= \inf \lset \Tr X + \sqrt{k} \sqrt{ \< R, R\>_{\sigma}}\bar - X \leq \rho - R \leq X \rset\\
  &= \sup \lset \Tr(Q-P) \rho \bar P + Q = \id,\; \Tr (Q-P)^2 \sigma \leq k ,\; P, Q \geq 0\rset\\
  &= \sup \lset \Tr(\id-2 P) \rho \bar \Tr (\id - 2P)^2 \sigma \leq k ,\; P \geq 0\rset,
\end{aligned}\end{equation}
where the second line is a duality argument following the same sequence of steps as above. This is the very same optimisation problem that we encountered in~\eqref{eq:lagdual}, showing the equivalence of the expressions.

The final claim that it suffices to optimise over classical--quantum operators follows because the dephasing $\Delta_X(\cdot) = \sum_x \proj{x} \cdot \proj{x}$ cannot change the optimal  value due to the fact that both $\rho_{XE}$ and $\id_X \otimes \sigma_E$ are invariant under it. For the optimisation over $R_{XE}$, this was already observed in~\cite[Theorem~14]{regula_2026}, and for the optimisation over $M_{XE}$ it can be seen that $\Tr M_{XE} \rho_{XE} = \Tr M_{XE} \Delta_X(\rho_{XE}) = \Tr \Delta_X(M_{XE}) \rho_{XE}$ but $\Delta_X(M_{XE})$ is always feasible for any feasible $M_{XE}$, since $\Tr [\Delta_X(M_{XE})]^2 [\id_X \otimes \sigma_E] \leq \Tr \Delta_X(M_{XE}^2) [\id_X \otimes \sigma_E] = \Tr M_{XE}^2 [\id_X \otimes \sigma_E]$ by Kadison's inequality~\cite[Theorem~2.3.2]{bhatia_2007}.
\end{proof}

From the above reformulation, a tight bound on the exponent of privacy amplification can be obtained straightforwardly, avoiding the difficulties in the proof of~\cite{regula_2026} that required a careful balancing of smoothing parameters. 
The result makes use of the variational form of measured R\'enyi divergences~\cite{berta_2017-1}, although here in fact we only need to lower bound measured R\'enyi, which only requires a classical H\"older's inequality. 
\begin{boxed}
\begin{corollary}\label{cor:err_exponent}
Under any 2*-universal family of hash functions, for all $\alpha \in (1,2]$ we have
\begin{equation}\begin{aligned}\label{eq:corollary_dupuis}
  \EE_f \frac12 \norm{\rho_{ZE}^f - \frac{\id_Z}{|\Z|} \otimes \rho_E}{1} \leq 4^{- \frac{\alpha-1}{\alpha}}\, \exp\left(- 
    \frac{\alpha-1}{\alpha} \left( H_\alpha^{\MM,\uparrow} (X|E)_\rho - \log |\Z| \right) \right),
\end{aligned}\end{equation}
recovering the result of~\cite[Theorem~16]{regula_2026}. Under 2-universal hash functions, the bound
\begin{equation}\begin{aligned}\label{eq:corollary_dupuis2}
  \EE_f \frac12 \norm{\rho_{ZE}^f - \frac{\id_Z}{|\Z|} \otimes \rho_E}{1} \leq 2^{- \frac{\alpha-1}{\alpha}}\, \exp\left(- 
    \frac{\alpha-1}{\alpha} \left( H_\alpha^{\MM,\uparrow} (X|E)_\rho - \log |\Z| \right) \right)
\end{aligned}\end{equation}
holds instead.
\end{corollary}
\end{boxed}
\begin{proof}
Recall that~\cite{berta_2017-1}
\begin{equation}\begin{aligned}
  D_{\alpha}^{\MM}(\rho \| \sigma) = \log \sup_{X \geq 0} \frac{\left(\Tr X \rho\right)^{\frac{\alpha}{\alpha-1}}}{\Tr X^{\frac{\alpha}{\alpha-1}}\sigma}.
\end{aligned}\end{equation}
Letting $M$ be any feasible POVM element in the variational form of $\hqr\pi_{k}(\rho\|\sigma)$ (Eq.~\eqref{eq:variational}), we then have
\begin{equation}\begin{aligned}
   D_{\alpha}^{\MM}(\rho_{XE} \| \id_X \otimes \sigma_E) &\geq \frac{\alpha}{\alpha-1} \log \Tr M_{XE} \rho_{XE} - \log \Tr M^{\frac{\alpha}{\alpha-1}}_{XE} [\id_X \otimes \sigma_E]\\
  &\geq \frac{\alpha}{\alpha-1}\log \Tr M_{XE} \rho_{XE} - \log \Tr M_{XE}^{2} [\id_X \otimes \sigma_E]\\
  &\geq \frac{\alpha}{\alpha-1}\log \Tr M_{XE} \rho_{XE} - \log k,
\end{aligned}\end{equation}
where we used that $\frac{\alpha}{\alpha-1} \geq 2$ and the assumption that $M$ is feasible for~\eqref{eq:variational}. Optimising over all feasible $M_{XE}$, and finally optimising over all $\sigma_E$ gives
\begin{equation}\begin{aligned}
  \log \inf_{\sigma_E} \hqr\pi_{|\Z|/4}  \!\left(  \rho_{XE} \middle\| \id_X \otimes \sigma_E \right) \leq \frac{\alpha-1}{\alpha} \left( -H_\alpha^{\MM,\uparrow} (X|E)_\rho + \log |\Z| - \log 4 \right).
\end{aligned}\end{equation}
and analogously for $k = |\Z|/2$. 
The proof is concluded by invoking Lemma~\ref{lem:lhl}.
\end{proof}
\begin{remark}
Relaxing the measured R\'enyi entropies $H_\alpha^{\MM,\uparrow} (X|E)_\rho$ in~\eqref{eq:corollary_dupuis} to sandwiched ones, $\wt{H}_\alpha^{\uparrow} (X|E)_\rho$, recovers the bound of~\cite[Theorem~8]{dupuis_2023}. The analogous bound for 2-universal hashes, i.e.\ the relaxation of~\eqref{eq:corollary_dupuis2}, was shown in~\cite[Corollary~3.5.6]{kamin_2026}. 
\end{remark}

\subsection{Asymptotics of the hypothesis testing bound}

To understand the tightness of the bound obtained from leftover hashing, we first ask: is the R\'enyi-based bound in Corollary~\ref{cor:err_exponent} tight as a bound on the exponent of the half-quadratic hypothesis testing problem that defines $\hqr\pi$? The answer is yes, and the asymptotic behaviour can be characterised exactly by establishing a connection with the strong converse exponent of standard quantum hypothesis testing~\cite{mosonyi_2015,hayashi_2016-1}.

\begin{tboxed}[white]{Strong converse exponent of half-quadratic hypothesis testing}
\begin{lemma}\label{lem:exponent_halfquad_testing}
For all quantum states
$\rho$ and $\sigma$ satisfying $\supp \rho \subseteq \supp \sigma$,
it holds that
  \begin{equation}\begin{aligned}
    \lim_{n\to\infty} -\frac1n \log \left[ \hqr\pi_{\exp(-n R)} (\rho^{\otimes n} \| \sigma^{\otimes n}) \right] = \sup_{\alpha \in (1,2]} \frac{\alpha-1}{\alpha} \left( R - \wt{D}_{\alpha}(\rho\|\sigma) \right),
  \end{aligned}\end{equation}
  and for classical--quantum states we have
  \begin{equation}\begin{aligned}
    \lim_{n\to\infty} -\frac1n \log \left[ \inf_{\sigma_{E^n}}\, \hqr\pi_{\exp(n R) } \!\left( \rho_{XE}^{\otimes n} \middle\| \id_{X^n} \otimes \sigma_{E^n} \right) \right] = \sup_{\alpha \in (1,2]} \frac{\alpha-1}{\alpha} \left( \wt{H}^{\uparrow}_{\alpha}(X|E)_\rho - R \right).
  \end{aligned}\end{equation}
\end{lemma}
\end{tboxed}
\begin{proof}
The converse direction (lower bound) follows from Corollary~\ref{cor:err_exponent}, as  $H_\alpha^{\MM,\uparrow} (X|E)_\rho \geq \wt{H}_\alpha^{\uparrow} (X|E)_\rho$. It remains to prove the achievability.  
To this end, let us recall once again the definitions
\begin{equation}\begin{aligned}
  \pi_\ve(\rho \| \sigma) &= \sup \lset \Tr M \rho \bar 0 \leq M \leq \id,\; \Tr M \sigma \leq \ve \rset,\\
  \hqr\pi_\ve (\rho \| \sigma) &= \sup \lset \Tr M \rho \bar 0 \leq M \leq \id,\; \Tr M^2 \sigma \leq \ve \rset.
\end{aligned}\end{equation}
It is easy to see that any POVM feasible for conventional hypothesis testing is feasible also for the half-quadratic one, i.e.\ $\hqr\pi_\ve (\rho \| \sigma) \geq \pi_\ve (\rho \| \sigma)$, because $M^2 \leq M$. However, we can say something stronger: for any POVM $M$ with $\Tr M \sigma \leq \ve$, $M' \coloneqq k M$ is also feasible for any $k \in (0,1]$, and furthermore $\Tr M'^2 \sigma \leq k^2 \Tr M \sigma \leq k^2 \ve$; hence,
\begin{equation}\begin{aligned}
  \hqr\pi_\ve(\rho \| \sigma) \geq k\, \pi_{\ve/k^2}(\rho \| \sigma).
\end{aligned}\end{equation}
Apply this now with the choice $\ve = \exp(-n R)$ and $k = \exp( - n \frac{ 1 - \lambda}{2} R )$ for some $\lambda \in [0,1]$. Then
\begin{equation}\begin{aligned}
  \hqr\pi_{\exp(-nR)}(\rho \| \sigma) \geq \exp\left(-n \frac{1-\lambda}{2} R\right) \pi_{\exp(-n \lambda R)} (\rho\|\sigma).
\end{aligned}\end{equation}
Applying this to i.i.d.\ copies and invoking the known strong converse exponent of hypothesis testing~\cite[Theorem~IV.10]{mosonyi_2015}, we get upon optimising $\lambda$ that
\begin{equation}\begin{aligned}
   \limsup_{n\to\infty} - \frac1n \log \hqr\pi_{\exp(-nR)}(\rho^{\otimes n}\| \sigma^{\otimes n}) &\leq \inf_{\lambda \in [0,1]} \left[\frac{1-\lambda}{2} R + \sup_{\alpha > 1} \frac{\alpha-1}{\alpha} \Big( \lambda R - \wt{D}_\alpha(\rho\|\sigma) \Big) \right]\\
   &= \sup_{\alpha > 1}\, \inf_{\lambda \in [0,1]} \left[ \left(\frac{1-\lambda}{2} + \frac{\alpha-1}{\alpha} \lambda \right) R - \frac{\alpha-1}{\alpha} \wt{D}_\alpha(\rho\|\sigma) \right]\\
   &=  \sup_{\alpha > 1} \left[ \left(\frac12\wedge\frac{\alpha-1}{\alpha}\right) R - \frac{\alpha-1}{\alpha} \wt{D}_{\alpha}(\rho\|\sigma) \right]\\
   &= \sup_{\alpha \in (1,2]} \frac{\alpha-1}{\alpha} \Big( R - \wt{D}_\alpha(\rho\|\sigma) \Big),
 \end{aligned}\end{equation} 
 where in the second line we used a minimax argument,  the applicability of which is perhaps easiest to see by reparametrising $\alpha \mapsto t^{-1}$ and writing
 \begin{equation}\begin{aligned}\label{eq:reparam}
   \sup_{\alpha > 1} \frac{\alpha-1}{\alpha} \Big( \lambda R - \wt{D}_\alpha(\rho\|\sigma) \Big) = \sup_{t \in (0,1)} \left[ (1-t) \lambda R - t \log \wt{Q}_{\frac{1}{t}}(\rho\|\sigma) \right]
 \end{aligned}\end{equation}
 where $\log \wt{Q}_\alpha = (\alpha - 1) \widetilde{D}_\alpha$; 
as $\alpha \mapsto \log \wt{Q}_{\alpha}$ is known to be convex~\cite[Lemma~III.3]{mosonyi_2015-1}, the function $t \mapsto t \log \wt{Q}_{\frac{1}{t}}$ is a perspective transform of a convex function and thus convex itself, making the overall expression on the RHS of~\eqref{eq:reparam} concave in $t$. 

The same holds for the conditional entropy $\wt{H}_{\alpha}^\uparrow$ using the analogous composite strong converse exponent of~\cite[Theorem~14]{hayashi_2016-1}, namely the fact that
\begin{equation}\begin{aligned}
  \lim_{n\to\infty} - \frac1n \log \left[ \inf_{\sigma_{E^n}} \pi_{\exp(n R) } \!\left(\rho_{XE}^{\otimes n}\middle\| \id_{X^n} \otimes\sigma_{E^n}\right) \right] = \sup_{\alpha > 1} \frac{\alpha-1}{\alpha} \left( \wt{H}_{\alpha}^{\uparrow}(X|E)_\rho - R \right);
\end{aligned}\end{equation} 
using this we can follow the proof steps above to give the stated result.
\end{proof}

\subsection{Tight one-shot converse and exponent under 4-universal hashing}
\label{subsec:tight_random_binning}


We now wish to ask whether leftover hashing gives an ensemble-tight bound on the error exponent of privacy amplification itself --- that is, whether the asymptotic exponential behaviour of half-quadratic hypothesis testing that we characterised in Lemma~\ref{lem:exponent_halfquad_testing} determines the asymptotics of universal randomness extraction.

However, even classically, 2-universality is not enough to get a matching converse.
Prior results in this direction~\cite{hayashi_2013,hayashi_2016-3} apply only to the particular case where the hash function is chosen independently at random among \emph{all} functions --- this corresponds to full mutual independence, or strong $k$-universality for all $k$, which is known as random binning in information theory literature or as strong universality$_\omega$ in the terminology of Wegman and Carter~\cite{wegman_1981}.\footnote{Here we point out for clarity that the works~\cite{hayashi_2013,hayashi_2016-3} made the rather non-standard choice of using the name `strongly universal$_2$' to refer to hash functions that are not merely strongly 2-universal, but strongly $k$-universal for all $k \in \NN$; the full mutual independence of the hash functions is explicitly used in the proofs there.}
 Furthermore, the asymptotic analysis has been shown to be tight only up to a critical rate: the bounds of Hayashi~\cite{hayashi_2013,hayashi_2016-3} are based on smooth min-entropy or information spectrum techniques, which naturally lead to an optimisation range of $\alpha \in (1, \infty)$, and thus they only match the known achievability results for rates large enough that the optimal $\alpha$ can be proven to satisfy~$\alpha \leq 2$.

We will give an improved converse bound that applies also to quantum side information, establishing for the first time the asymptotic tightness of the exponent derived from the leftover hash lemma. 
Importantly, our result is tight up to a constant, showing that the leftover hash lemma gives an optimal achievability not only asymptotically, but even as a one-shot bound on the error.
We also do not need to assume full independence of the hash functions: although we need to go beyond 2-universality, we only make use of four-wise independence of the hashes.
Our asymptotic result is an improvement even in the classical case, where it removes the need for a critical rate and shows the exponent to be tightly governed by R\'enyi conditional information in the range of $\alpha \in (1,2]$.

To unburden the notation, let us define a shorthand for the success probability of half-quadratic hypothesis testing optimised over states on the system $E$ as
\begin{equation}\begin{aligned}
  \hqr\pi_{k}(X|E)_\rho &\coloneqq \inf_{\sigma_E} \,\hqr\pi_{ k / |\X|}  \!\left( \rho_{XE} \middle\| \frac{\id_X}{|\X|} \otimes \sigma_E \right)\\
  &\hphantom{:}= \! \inf_{\ve \in [0,1)}  \left[ \ve + \sqrt{k}\, \exp\!\left(-\frac12 \Hs{2}{\ve}{M}(X|E)_\rho\right) \right]\\
  &\hphantom{:}= \inf \lset \norm{Q_{XE}}{1} + \sqrt{k} \, \inf_{\sigma_E} \norm{R_{XE}}{\id_X \otimes \sigma_E} \bar \rho_{XE} = Q_{XE} + R_{XE}, \; Q_{XE} \geq 0 \rset\\
  &\hphantom{:}= \sup \lset \Tr M_{XE} \rho_{XE} \bar 0 \leq M_{XE} \leq \id,\; \norm{\Tr_X M_{XE}^2}{\infty} \leq k \rset,
\end{aligned}\end{equation}
where the equalities follow from Lemma~\ref{lem:lhl}.
Recall that the leftover hash lemma tells us that
\begin{equation}\begin{aligned}
  \EE_f \frac12 \norm{\rho_{ZE}^f - \frac{\id_Z}{|\Z|} \otimes \rho_E}{1} \leq \hqr\pi_{|\Z|/4}(X|E)_\rho 
\end{aligned}\end{equation}
for any 2\*-universal family of hash functions. We are then interested in the opposite inequality.

\begin{tboxed}[filled]{Converse for leftover hash lemma}
\begin{theorem}\label{thm:reallytight}
For any 4*-universal family of hash functions $f \colon \X \to \Z$ with $|\X| \geq 4$ and $|\Z| \geq 2$, the achievability bound of the leftover hash lemma is tight up to a constant:
\begin{equation}\begin{aligned}
\EE_f \frac12 \norm{\rho_{ZE}^f - \frac{\id_Z}{|\Z|} \otimes \rho_E}{1} \geq \frac{1}{3}\, \hqr\pi_{|\Z|/4}(X|E)_\rho.
\end{aligned}\end{equation}

As a consequence, for the class of 4*-universal hash functions $\univstar{4}$ or any subset thereof, we have
\begin{equation}\begin{aligned}
   \lim_{n\to\infty} - \frac1n \log \Delta_{\min,\univstars{4}}\! \left(\rho^{\otimes n}_{XE},\, \exp(nR)\right) = \sup_{\alpha \in (1,2]} \frac{\alpha-1}{\alpha} \left( \wt{H}^{\uparrow}_{\alpha}(X|E)_\rho - R \right)
\end{aligned}\end{equation}
for all rates $R > 0$, with the exponent achieved by any 2-universal hash family.
\end{theorem}
\end{tboxed} 

Before proceeding with the proof, we discuss some of the rather interesting connections that emerge here. In the simple case of no side information, one can easily see that we are dealing with a form of an infimal convolution: the optimisation reduces to
\begin{equation}\begin{aligned}\label{eq:peere}
  \hqr\pi_{|\Z|/4}(X)_p = \inf_{\ve \in [0,1)} \left[ \ve +  \sqrt{ \frac{|\Z|}{4} \, Q_2^{\ve,\MM} \left( p_{X} \| \id_X \right) } \right] =  \inf_{p = q + r} \left[ \norm{q}{1} + \sqrt{\frac{|\Z|}{4}} \norm{r}{2} \right].
\end{aligned}\end{equation} 
This is evocative of the fact that we are interpolating between $\ell_1$-type and $\ell_2$-type norms, but --- unlike in results that use complex interpolation techniques --- in an additive way.\footnote{In quantum information, the same quantity also appeared in the study of coherence and entanglement distillation under the name of $m$-distillation norm~\cite{regula_2017,regula_2019}.}  
More generally, the quantity on the right-hand side above is known as the Peetre $K$-functional,  which one can define in a general manner as
\begin{equation}\begin{aligned}
  K(x, t; X_0, X_1) \coloneqq \inf_{x = y + z} \left[ \norm{y}{X_0} + t \norm{z}{X_1} \right]
\end{aligned}\end{equation}
for some suitable spaces $X_0$, $X_1$. Such functionals indeed originate in \emph{real} interpolation theory~\cite{holmstedt_1970}, but a particularly important application of them is in the theory of symmetric independent random variables: the asymptotic behaviour of many symmetric random processes can be characterised using celebrated inequalities due to Khintchine, Rosenthal~\cite{rosenthal_1970}, and more general variants thereof~\cite{johnson_1989,montgomery-smith_1990,hitczenko_1993,hitczenko_1994}, whose optimal forms are indeed given by suitable $K$-functionals~\cite{johnson_1989,hitczenko_1994}. 

Since our Theorem~\ref{thm:reallytight} is concerned with a similar $K$-functional, it is natural to seek connections with non-commutative generalisations of the theory of symmetric random variables. However, several issues prevent the direct application of known results. Standard non-commutative Khintchine inequalities are limited to specific choices of independent random variables such as Rademacher or Gaussian ones~\cite{lust-piquard_1991,pisier_1997,pisier_2009} which does not suffice to establish the tight estimate needed in our setting. While more general results are known, they come with other limitations: they either apply only to $L_p$ spaces with $p \in (0,1) \cup (1,\infty)$ and explicitly exclude the endpoint case of $L_1$~\cite{junge_2008} --- which is precisely the trace norm that we need to bound in our case --- or they make stronger assumptions about the independence or identicality of the random variables~\cite{junge_2010}, which again makes them inapplicable to the problem at hand.
We will then take a different route and adapt the proof approach of Haagerup and Musat~\cite{haagerup_2007}, originally introduced to give a refined derivation of non-commutative Khintchine inequalities. Their insightful and careful techniques are straightforward to generalise beyond the specific cases studied there, and indeed will allow us to establish precise estimates for trace distance, with the added benefit of only requiring four-wise independence.

For the sake of readability, we will avoid invoking operator-space notation and terminology. This will help us accentuate the fact that the derivation --- while modestly cumbersome --- is elementary in nature.
Later in Section~\ref{subsec:connections} we will discuss in more detail how the quantities studied in our work are directly connected to the operator norms that frequently find use in contexts related to non-commutative Khintchine-type inequalities.

\begin{remark}[Proof idea]
For any $z \in \Z$, the occupancy of the $z$th bin is characterised by the random variables $\delta_{f(x),z}$. Setting $p \coloneqq \frac{1}{|\Z|}$, let us define a shifted variant of these random variables as
\begin{equation}\begin{aligned}
  B_x^{(z)} \coloneqq \delta_{f(x),z} - p.
\end{aligned}\end{equation}
For a general CQ state $\rho_{XE} = \sum_x \proj{x} \otimes \rho_{E,x}$, we can now write
\begin{equation}\begin{aligned}
  \EE_f \frac12 \norm{ \rho^f_{ZE} - \frac{\id_Z}{|\Z|} \otimes \rho_E }{1} &= \EE_f \sum_z \frac12 \norm{\rho^f_{E,z} - \frac{1}{|\Z|} \rho_E}{1}\\
  &= \EE_f \sum_z \frac12 \norm{ \sum_x B^{(z)}_x \, \rho_{E,x}}{1},
\end{aligned}\end{equation}
and our aim is to lower bound this quantity in terms of $\hqr\pi_{|\Z|/4}(X|E)_\rho$. The key idea will be to employ the dual optimisation form of the latter problem, namely
\begin{equation}\begin{aligned}\label{eq:trianglenorm_dual}
  \hqr\pi_{|\Z|/4}(X|E)_\rho &= \sup \lset \sum_x \Tr M_{E,x} \rho_{E,x} \bar 0 \leq M_{E,x} \leq \id \; \forall x,\; \sum_x M_{E,x}^2 \leq \frac{1}{4p} \id \rset.
\end{aligned}\end{equation}
Consider that if we define $\widebar{M}_{E,z} \coloneqq \sum_{x} B^{(z)}_x M_{E,x}$, then
\begin{equation}\begin{aligned}
  \EE_f \Tr \left[ \left( \rho^f_{E,z} - \frac{1}{|\Z|} \rho_E \right) \widebar{M}_{E,z} \right] &= \EE_f \left[ \sum_{x,x'} B^{(z)}_x B^{(z)}_{x'} \Tr \rho_{E,x} M_{E,x'} \right]\\
  &= p (1-p) \sum_x \Tr \rho_{E,x} M_{E,x}\\
  &\geq p \frac12 \sum_x \Tr \rho_{E,x} M_{E,x},
\end{aligned}\end{equation}
where we used that $\EE_f [ B^{(z)}_x B^{(z)}_{x'} ] = p(1-p)$ when $x = x'$ and 0 otherwise, and that $|\Z|\geq 2$.
What this means is that, \emph{if} it were true that $\norm{\widebar{M}_{E,z}}{\infty} \leq 1$ where $(M_{E,x})_x$ are the optimal operators in~\eqref{eq:trianglenorm_dual}, then from the dual expression $\norm{X}{1} = \max \lset \Tr X W \bar \norm{W}{\infty} \leq 1 \rset$ we would immediately get
\begin{equation}\begin{aligned}
  \EE_f \frac12 \norm{ \rho^f_{ZE} - \frac{\id_Z}{|\Z|} \otimes \rho_E  }{1} &\stackrel{?}{\geq} \sum_z \EE_f \frac12 \Tr \left[ \left( \rho^f_{E,z} - \frac{1}{|\Z|} \rho_E \right) \widebar{M}_{E,z} \right]\\ 
  &\geq |\Z|\, p \frac{1}{4} \sum_x \Tr \rho_{E,x} M_{E,x}\\
  &=\frac14 \,\hqr\pi_{|\Z|/4}(X|E)_\rho,
\end{aligned}\end{equation}
which is (up to a constant) exactly the bound that we wish to establish. But in general $\smash{\norm{\widebar{M}_{E,z}}{\infty}} \not\leq 1$, and the main difficulty in the proof is precisely to massage the operators $\widebar{M}_{E,z}$ into a form whose norm we can control.  In order to do this, we will follow Haagerup and Musat~\cite{haagerup_2007} in constructing a truncated version of the operator $\widebar{M}_{E,z}$ that can be shown to remain bounded `in expectation' provided that we can control some of its higher moments, which is precisely where the assumption of 4\*-universality is needed.
\end{remark}

The properties of the random variables $(B^{(z)}_x)_x$ and the operators  $\widebar{M}_{E,z}$ built from them are determined by the assumptions on the given family of hash functions. In particular, if the hash functions were strongly 4-universal, the random variables $(B^{(z)}_x)_x$ would be four-wise mutually independent, which would greatly simplify the analysis of their higher moments. 
We begin by clarifying that, through an additional randomisation, the assumption of 4\*-universality is enough to allow for the same independence property to hold --- effectively meaning that the proof can proceed in a way similar to assuming strong 4-universality.

\begin{lemma}[On 4\*-universality and mutual independence]\label{lem:bernoulli}
Randomising the choice of bins for any 4\*-universal hash family can, without affecting the error of randomness extraction, mimic the four-wise independence ensured by strong 4-universality: specifically, without loss of generality, the random variables $(B^{(z)}_x)_x$ with $B^{(z)}_x = \delta_{f(x),z} - p$ can be assumed to be four-wise independent, centred $\operatorname{Bernoulli}(p)$ random variables with $\EE[B^{(z)}_x] = 0$.
\end{lemma}
\begin{proof}
Let $u$ be uniform on $\Z$ independently of $f$. For any distinct $x_1,\ldots,x_j$ with $1\leq j\leq4$, we then have
\begin{align}
\EE_{f,u}\prod_{i=1}^j \delta_{f(x_i), u} = \frac{1}{|\Z|} \Pr_f\Big(f(x_1)=\cdots=f(x_j)\Big) = p^j
\end{align}
with $p = \frac{1}{|\Z|}$ as before. 
For $j=1$, the last equality here follows from the uniformity of $u$; for $2\leq j\leq4$, it follows from the assumed 4*-universality. These product moments determine the joint law of any four binary indicators. Hence, the indicators $\delta_{f(x_i), u}$ are four-wise independent Bernoulli random variables under the joint randomness $(f,u)$. 
Now, the trace norm decomposes over the classical output bins, which gives
\begin{align}
\EE_f\frac12
\norm{\rho^f_{ZE}-\frac{\id_Z}{|\Z|}\otimes\rho_E}{1}
= \frac{|\Z|}{2}\EE_{f,u}
\norm{\sum_x\left(\delta_{f(x),u}-\frac{1}{|\Z|}\right)\rho_{E,x}}{1}
\end{align}
and thus shows that four-wise independence can be assumed without loss of generality, up to identifying $B^{(z)}_x$ with the additionally randomised random variables $\delta_{f(x), u} - p$.
\end{proof}

We proceed with a lemma that improves on a similar estimate shown in~\cite[Proposition~2.12]{haagerup_2007}. For reasons that will become clear shortly, we need an upper bound on the fourth moments of $\ol{M}$ that scales as $p(1-p)$, and following the derivation in~\cite{haagerup_2007} would not have been sufficient to obtain this scaling. 
The key difference for us is that standard Khintchine inequalities only depend on the norm $\norm{\sum_x M_x^2}{\infty}$, but in our setting we can also make use of the fact that each $\norm{M_x}{\infty}$ is bounded, which allows us to trade off between two different norms, giving a more favourable scaling of the bound. 

Below we will suppress the bin index and use the notation $B_x$ to denote $\delta_{f(x),z} - p$ or $\delta_{f(x),u} - p$, as the analysis will not depend on the specific choice of the bin. 

\begin{lemma}[Fourth-moment estimate]\label{lem:4thmoment}
Consider any collection of Hermitian operators $(M_x)_x$, and let $\widebar{M} \coloneqq \sum_x B_x M_x$ where $(B_x)_x$ are four-wise independent centred $\operatorname{Bernoulli}(p)$ random variables. Then
\begin{equation}\begin{aligned}
  \EE \widebar{M}^2 &= p(1-p) \sum_{x} M_{x}^2,\\
  \EE \widebar{M}^4 &\leq 
  \frac{7}{4}\, p(1-p)^3 \, \max \left\{ \max_x \norm{M_x}{\infty}^2, \; 4p \norm{\sum_x M_x^2}{\infty} \right\}\,  \sum_{x} M_{x}^2.
\end{aligned}\end{equation}
\end{lemma}
\begin{proof}
By assumption, the random variables $(B_x)_x$ satisfy that 
\begin{equation}\begin{aligned}\label{eq:pairwise_exp}
  \EE [ B_x B_{x'} ] = \begin{cases} p(1-p) & \text{ if } x = x'\\
  0 & \text{ otherwise,}\end{cases}
\end{aligned}\end{equation}
which immediately gives $\EE \widebar{M}^2 = \sum_{x} p(1-p)\, M_{x}^2$. 
Using the four-wise independence of the random variables and the fact that $\EE_f [B_x] = 0$, we also deduce that
\begin{equation}\begin{aligned}
  \EE [ B_{x_1} B_{x_2} B_{x_3} B_{x_4} ] =  
      \begin{cases}
         p(1-p)^4 + (1-p)p^4 & \text{ if } x_i = x_j \text{ for all } i,j\\
         p^2(1-p)^2 & \text{ if } x_i = x_j \neq x_k = x_l \text{ for any pairs } (i,j) \text{ and } (k,l)\\
         0 & \text{ otherwise.}
       \end{cases}
\end{aligned}\end{equation}
This implies that
\begin{equation}\begin{aligned}\label{eq:thisimplies}
  \EE \widebar{M}^4 &= \left[p(1-p)^4 + (1-p)p^4\right] \sum_{x} M_{x}^4 + p^2 (1-p)^2  \sum_{x \neq x'} \left( M_{x}^2 M_{x'}^2 + M_{x} M_{x'}^2 M_{x} + M_{x} M_{x'} M_{x} M_{x'} \right).
\end{aligned}\end{equation}
Notice now that
\begin{equation}\begin{aligned}
  \sum_{x \neq x'} M_{x} M_{x'} M_{x} M_{x'} &= \sum_{x < x'} \left( M_{x} M_{x'} M_{x} M_{x'} + M_{x'} M_{x} M_{x'} M_{x}\right)\\
  &\leq \sum_{x < x'} M_{x} M_{x'}^2 M_{x} + M_{x'} M_{x}^2 M_{x'}\\
  &= \sum_{x \neq x'} M_{x} M_{x'}^2 M_{x},
\end{aligned}\end{equation}
where the inequality follows since for any Hermitian $A$ and $B$ we have
\begin{equation}\begin{aligned}
  0 \leq ( A B - B A)^\dagger ( A B - B A ) = A B^2 A + B A^2 B - A B A B - B A B A.
\end{aligned}\end{equation}
The second term in Eq.~\eqref{eq:thisimplies} can then be upper bounded as
\begin{align}
\sum_{x \neq x'} \left( M_{x}^2 M_{x'}^2 + M_{x} M_{x'}^2 M_{x} + M_{x} M_{x'} M_{x} M_{x'} \right) &\leq  \sum_{x \neq x'}  \left( M_{x}^2 M_{x'}^2 + 2 M_{x} M_{x'}^2 M_{x} \right) \\
  &=  \sum_{x}  \left( M_{x}^2 \left[ \sum_{x' \neq x} M_{x'}^2 \right] + 2 M_{x} \left[ \sum_{x' \neq x} M_{x'}^2 \right] M_{x} \right) \nonumber\\
  &=  \sum_{x}  \left( M_{x}^2 \left[ \sum_{x'} M_{x'}^2 - M_{x}^2 \right] + 2 M_{x} \left[ \sum_{x' } M_{x'}^2 - M_{x}^2\right] M_{x} \right) \nonumber\\
  &=  \left( \sum_{x} M_{x}^2 \right)^2  +  2  \sum_x M_{x} \left[ \sum_{x' } M_{x'}^2 \right] M_{x} - 3 \sum_x M_{x}^4 \nonumber\\
  &\leq  \norm{\sum_x M_x^2}{\infty}\sum_x M_{x}^2  +  2 \norm{\sum_x M_x^2}{\infty} \sum_x M_{x} ^2 - 3 \sum_x M_{x}^4. \nonumber
\end{align}
Plugging this back into~\eqref{eq:thisimplies} gives
\begin{equation}\begin{aligned}
  \EE \widebar{M}^4 &\leq \left[p(1-p)^4 + (1-p)p^4 - 3 p^2 (1-p)^2 \right] \sum_{x} M_{x}^4 + 3 p^2 (1-p)^2 \norm{\sum_x M_x^2}{\infty} \sum_x M_{x}^2 \\
  &= p(1-p) \left( \left[ 1 - 6p (1-p)  \right] \sum_{x} M_{x}^4 + 3 p (1-p) \norm{\sum_x M_x^2}{\infty} \sum_x M_{x}^2 \right).
\end{aligned}\end{equation}
From this we can already extract an upper bound in terms of a common operator $\sum_x M_{x}^2$. Keeping a careful track of all terms in order to optimise the overall multiplicative scaling factor, we obtain the announced estimate as
\begin{align}
    \EE \widebar{M}^4 &\textleq{(i)} p(1-p) \left( \left(1- 2 p\right)^2   \sum_{x} M_{x}^4 + 3 p (1-p) \norm{\sum_x M_x^2}{\infty} \sum_x M_{x}^2 \right) \nonumber\\
  &\leq p(1-p) \left(  \left(1- 2 p\right)^2 \sum_{x} \norm{M_x}{\infty}^2 M_{x}^2 + 3 p (1-p) \norm{\sum_x M_x^2}{\infty} \sum_x M_{x}^2\right) \raisetag{4\baselineskip} \\
  &\leq p(1-p) \left(  \left(1- 2 p\right)^2 \max_x \norm{M_x}{\infty}^2 \sum_{x} M_{x}^2 + 3 p (1-p) \norm{\sum_x M_x^2}{\infty} \sum_x M_{x}^2\right) \nonumber \\
  &\leq p(1-p) \left(  \left(1- 2p\right)^2 + \frac34(1-p) \right) \, \max \left\{ \max_x \norm{M_x}{\infty}^2, \; 4p \norm{\sum_x M_x^2}{\infty} \right\}\, \sum_x M_{x}^2, \nonumber\\
  &\textleq{(ii)} p(1-p) \frac74 (1-p)^2 \, \max \left\{ \max_x \norm{M_x}{\infty}^2, \; 4p \norm{\sum_x M_x^2}{\infty} \right\}\, \sum_x M_{x}^2, \nonumber
\end{align}
where in (i) we simply bounded $1 - 6 p(1-p) \leq 1 - 4 p (1-p) = (1-2p)^2$, and in (ii) we observed that the assumption $|\Z| \geq 2$ entails that $p \leq \frac12 \leq (1-p)$ and hence
\begin{equation}\begin{aligned}
  \big((1-p)-p\big)^2 + \frac34 (1-p) &\leq (1-p) \left( (1-p)- 2 p + p + \frac34 \right) = (1-p) \left( \frac74 (1-p) - \frac14 p \right).
\end{aligned}\vspace*{-\baselineskip}\end{equation}
\end{proof}

The fourth-moment estimate can now be used to establish the following result, which adapts the operator truncation idea used in~\cite[Lemma~2.6]{haagerup_2007} (although the specific construction appeared even earlier e.g.\ in~\cite{kaijser_1983}). 

\begin{lemma}[Approximation by normalised operators]\label{lem:truncation}
Consider any Hermitian operators $(M_x)_x$
with
\begin{equation}\begin{aligned}
  \norm{M_x}{\infty} &\leq 1 \quad \forall x\\
   \norm{\sum_x M_x^2}{\infty} &\leq \frac{1}{4p}
\end{aligned}\end{equation}
and let $\ol{M} = \sum_x B_x M_x$. 
Then there exists a function $h: \RR \to \RR$ such that
\begin{equation}\begin{aligned}
    \norm{\ol{M} - h(\ol{M})}{\infty} &\leq \frac34 (1-p)
\end{aligned}\end{equation}
and such that the operators $K_x \coloneqq \frac{2}{p(1-p)} \EE \left[B_x h(\ol{M})\right]$ satisfy
\begin{equation}\begin{aligned}
  \norm{K_x}{\infty} &\leq 1 \quad \forall x,\\
  \norm{\sum_x K_x^2}{\infty} &\leq \frac{1}{4p}.
\end{aligned}\end{equation}
\end{lemma}

\begin{proof}
For some constant $c > 0$, consider the scalar function
\begin{equation}\begin{aligned}
  g(t) \coloneqq \begin{cases} c & \text{ if } t > c \\ t & \text{ if } |t| \leq c \\ -c & \text{ if } t < - c. \end{cases}
\end{aligned}\end{equation}
The function has two key properties. Clearly, $|g(t)| \leq c$ for all $t \in \RR$, so $\norm{g(X)}{\infty} \leq c$  for any Hermitian $X$. 
What will be important is that we can also bound the function
\begin{equation}\begin{aligned}
  h(t) \coloneqq t - g(t) = 
  \operatorname{sgn}(t)\, (|t| - c)_+,
\end{aligned}\end{equation}
and hence the norm of $X - g(X)$, due to the fact that
\begin{equation}\begin{aligned}\label{eq:gt}
  \left| h(t) \right| \leq \frac{1}{4c} t^2 \quad \forall t \in \RR.
\end{aligned}\end{equation}
This is immediate for $|t| \leq c$, and for $|t| > c$ is easily seen by writing
\begin{equation}\begin{aligned}
  \frac{c(|t|-c)}{t^2} = \frac{c}{|t|} \left( 1 - \frac{c}{|t|} \right) \leq \frac14.
\end{aligned}\end{equation}

Recalling the definition $\ol{M} = \sum_x B_x M_x$ and noting that $\EE [ B_x \ol{M} ] = p(1-p) M_x$, we now consider
\begin{equation}\begin{aligned}
  H_x \coloneqq \EE_f \!\left[ B_x \, h\!\left(\ol{M}\right) \right] = \EE_f \!\left[ B_x \!\left(\ol{M} - g(\ol{M})\right) \right].
\end{aligned}\end{equation}
To estimate the norm of these operators, observe first that clearly $- |B_x| \, |h(\ol{M})| \leq B_x \, h(\ol{M}) \leq |B_x| \, |h(\ol{M})|$. 
As $|B_x|$ is four-wise independent of $\{B_{x'}\}_{x' \neq x}$, we can compute 
\begin{align}
 \EE \left|B_x\right| B_{x'} B_{x''} &= \begin{cases} p (1-p)^3 + (1-p) p^3 & \text{ if } x = x' = x''\\ p \left[ p(1-p)^3 + (1-p)^2 p^2\right] + (1-p)\left[p^2 (1-p)^2 + (1-p)p^3 \right] & \text{ if } x \neq x' = x'' \\ 0 & \text{ if } x' \neq x''  \end{cases} \nonumber\\
 &= \begin{cases} p (1-p) \left[ p^2 + (1-p)^2 \right] & \text{ if } x = x' = x''\\ 2 p^2 (1-p)^2 & \text{ if } x \neq x' = x'' \\ 0 & \text{ if } x' \neq x''. \end{cases}
\end{align}
Using~\eqref{eq:gt} then gives
\begin{equation}\begin{aligned}
  \norm{H_x}{\infty} &\leq \norm{\EE\! \left[|B_x| \, |h(\ol{M})|\right]}{\infty}\\
  &\leq \frac{1}{4c} \norm{ \EE |B_x| \ol{M}^2 }{\infty} \\
  &= \frac{1}{4c}  \norm{\sum_{x',x''} \EE \left[|B_x|\, B_{x'} B_{x''}\right] M_{x'} M_{x''} }{\infty}\\
  &= \frac{1}{4c}  \norm{ p (1-p) \left[ p^2 + (1-p)^2 \right] M_x^2 + 2 p^2 (1-p)^2 \sum_{x' \neq x} M_{x'}^2 }{\infty}\\
  &= \frac{p(1-p)}{4c} \norm{ (2p-1)^2\, M_x^2 + 2 p (1-p) \sum_x M_x^2 }{\infty}.
\end{aligned}\end{equation}
Recalling the assumptions on $M_x$, we obtain
\begin{equation}\begin{aligned}\label{eq:estimate1}
  \norm{H_x}{\infty}  &\leq \frac{p(1-p)}{4c} \left[ (2p-1)^2 \norm{M_x}{\infty} + 2p(1-p) \norm{\sum_x M_x^2}{\infty}\right]\\
  &\leq \frac{p(1-p)}{4c} \left[ (2p-1)^2 + \frac12(1-p)\right]\\
  &\leq \frac{p(1-p)}{2} \, \frac{3}{4c} (1-p) ,
\end{aligned}\end{equation}
where the last line uses the fact that $(1-p) - (2p-1)^2 = p (3-4p) \geq p \geq 0$ since $|\Z| \geq 2$.

We will now bound the norm of $\sum_x H_x^2$. 
Since (recalling~\eqref{eq:pairwise_exp})
\begin{equation}\begin{aligned}
  \sum_x H_x^2 &= \frac{1}{p(1-p)} \sum_{x,x'} \EE [B_x B_{x'}] H_x H_{x'}\\
  &= \frac{1}{p(1-p)} \EE \left( \sum_{x} B_x H_x \right)^2,
\end{aligned}\end{equation}
we have that
\begin{align}
  0 &\leq \EE \left( p(1-p)\, h\!\left(\ol{M}\right) - \sum_x B_x H_x \right)^2 \nonumber\\
  &= p^2(1-p)^2\, \EE h\!\left(\ol{M}\right)^2 - p(1-p) \sum_x \EE\!\left[ h\!\left(\ol{M}\right) B_x\right] H_x - p(1-p) \sum_x H_x \EE\!\left[ B_x h\!\left(\ol{M}\right)\right] + \EE \left(\sum_x B_x H_x\right)^2 \nonumber\\
  &= p(1-p) \left[ p(1-p)\, \EE h\!\left(\ol{M}\right)^2 - \sum_x H_x^2 - \sum_x H_x^2 + \sum_x H_x^2 \right] \nonumber\\
  &= p(1-p) \left[ p(1-p)\, \EE h\!\left(\ol{M}\right)^2 - \sum_x H_x^2\right]
\end{align}
This allows us to use the fourth-moment estimate in Lemma~\ref{lem:4thmoment} to get
\begin{equation}\begin{aligned}\label{eq:estimate2}
  \sum_x H_x^2 &\leq p(1-p)\, \EE h\!\left(\ol{M}\right)^2\\
  &\textleq{(i)} \frac{p(1-p)}{16 c^2} \EE \ol{M}^4\\
  &\textleq{(ii)} \frac{p(1-p)}{16 c^2} \frac{7}{4} p(1-p)^3 \, \sum_x M_x^2\\
 &\textleq{(iii)} \frac{p^2(1-p)^2}{4} \,\frac{7 (1-p)^2}{16 c^2}\, \frac{1}{4p} \id,
\end{aligned}\end{equation}
with (i) by~\eqref{eq:gt}, (ii) by Lemma~\ref{lem:4thmoment} combined with the assumptions on $M_x$, and (iii) again using the hypothesis $\sum_x M_x^2 \leq \frac{1}{4p} \id$.

Choosing
\begin{equation}\begin{aligned}
  c =\frac34 (1-p) \quad\Rightarrow\quad  c^2 \geq\frac{7}{16}(1-p)^2 
\end{aligned}\end{equation}
and rescaling $H_x$ as
\begin{equation}\begin{aligned}
  K_x \coloneqq \frac{2}{p(1-p)} H_x,
\end{aligned}\end{equation}
the estimates in~\eqref{eq:estimate1} and~\eqref{eq:estimate2} tell us that $K_x \leq \id$ for all $x$ and that $\sum_x K_x^2 \leq \frac{1}{4p} \id$, which is what was to be shown.
\end{proof}

Armed with the two lemmas above, we are ready to establish the desired bound.

\begin{proof}[Proof of Theorem~\ref{thm:reallytight}]
Recall from Lemma~\ref{lem:lhl} that in the variational form of $\hqr\pi_{|\Z|/4}(X|E)_\rho$ it suffices to consider classical--quantum operators $M_{XE} = \sum_x \proj{x} \otimes M_{E,x}$, and that the assumption of positivity can be relaxed without loss of generality. This altogether gives the equivalent dual form
\begin{equation}\begin{aligned}\label{eq:trianglenorm_dual2}
 \hqr\pi_{|\Z|/4}(X|E)_\rho &= \sup \lset \sum_x \Tr M_{E,x} \rho_{E,x} \bar \norm{ M_{E,x} }{\infty} \leq 1 \; \forall x,\; \norm{\sum_x M_{E,x}^2}{\infty} \leq \frac{1}{4p} \rset.
\end{aligned}\end{equation}
Let $(M_{E,x})_x$ be an optimal solution for this optimisation. For any $z \in \Z$, consider the random variables $B^{(z)}_x = \delta_{f(x),z} - p$, and let us for simplicity assume that they are four-wise independent --- this can always be ensured through an additional averaging as in Lemma~\ref{lem:bernoulli}. Since any feasible $(M_{E,x})_x$  satisfy the assumptions of Lemma~\ref{lem:truncation}, we know that $\ol{M}_E \coloneqq \sum_x B^{(z)}_x M_{E,x} $ satisfies $\norm{ \frac{4}{3(1-p)} \, g(\ol{M}_E)}{\infty} \leq 1$ for the function $g(t) = t - h(t)$ as defined in the lemma. Then
\begin{align}
  \frac12 \EE_f \norm{ \rho^f_{E,z} - \frac{1}{|\Z|} \rho_E}{1} &=  \frac12 \EE \norm{\sum_x B^{(z)}_x \rho_{E,x}}{1} \nonumber\\
  &\geq \frac12 \frac{4}{3(1-p)} \EE \Tr \left[ \sum_x B^{(z)}_x \rho_{E,x} \, g(\ol{M}_E) \right] \nonumber\\
    &=  \frac{2}{3(1-p)} \EE \Tr \left[ \sum_x B^{(z)}_x \rho_{E,x} \left(\ol{M}_E - h(\ol{M}_E)\right) \right]\\
  &= \frac{2 p}{3} \left( \sum_x \frac{1}{p(1-p)} \Tr \rho_{E,x} \EE\left[B^{(z)}_x \ol{M}_E \right]- \frac{1}{p(1-p)} \sum_x \Tr \rho_{E_x} \EE\left[ B^{(z)}_x h(\ol{M}_E) \right] \right) \nonumber\\
  &=  \frac{2 p}{3} \left( \sum_x \Tr \rho_{E,x} M_{E,x} - \frac12 \sum_x \Tr \rho_{E_x} K_{E,x} \right). \nonumber
\end{align}
Recall now that we chose $M_{E,x}$ so that $\sum_x \Tr \rho_{E,x} M_{E,x} = \hqr\pi_{|\Z|/4}(X|E)_\rho$. Furthermore, since Lemma~\ref{lem:truncation} tells us that $(K_{E,x})$ are also feasible for the same optimisation problem in~\eqref{eq:trianglenorm_dual2}, we have $\sum_x \Tr \rho_{E,x} K_{E,x} \leq \hqr\pi_{|\Z|/4}(X|E)_\rho$. Thus
\begin{equation}\begin{aligned}
  \frac12 \EE_f \norm{ \rho^f_{E,z} - \frac{1}{|\Z|} \rho_E}{1} &\geq \frac{2 p}{3} \frac12 \,\hqr\pi_{|\Z|/4}(X|E)_\rho.
\end{aligned}\end{equation}
Applying the argument to each bin $z \in \Z$ gives
\begin{equation}\begin{aligned}
 \EE_f \frac12 \norm{ \rho^f_{ZE} - \frac{\id_Z}{|\Z|} \otimes \rho_E }{1} &= \sum_z \frac12 \EE_f\norm{ \rho^f_{E,z} - \frac{1}{|\Z|} \rho_E}{1}\\
 &\geq |\Z|\, \frac{p}{3} \,\hqr\pi_{|\Z|/4}(X|E)_\rho
\end{aligned}\end{equation}
and thus concludes the proof.
\end{proof}

\subsection{Connections with operator norms}\label{subsec:connections}

Let us clarify how the quantities studied in our work are related to the operator norms employed in prior works on non-commutative probability theory and functional analysis.
Related links were observed also in the work~\cite{berta_2017-2}, which studied cryptographic notions of extractors and made use of norm interpolation spaces. Very recently, interpolation functionals and non-commutative Rademacher inequalities also found conceptually similar applications in the works~\cite{li_2026,li_2026-1}, which however focused on other security criteria based on sandwiched R\'enyi divergences. 
Here, we managed to obtain tight bounds for the operationally relevant trace distance metric through the connections we established with the half-quadratic hypothesis testing problem based on the smooth measured conditional R\'enyi entropy of order 2 --- this, in fact, precisely corresponds to a known type of an operator-space interpolation norm.

One can imagine a CQ operator $A_{XE} = \sum_x \proj{x} \otimes A_{E,x}$ as a vector of operators $(A_{E,x})_x$, where each $A_{E,x}$ lives in the standard space of trace-class operators endowed with the Schatten norm $\|\cdot\|_1$. 
There are then two natural ways to define a norm on such a space, depending on whether we regard $(A_{E,x})_x$ as a row vector or column vector: they are the \deff{row norm} $\norm{\cdot}{\row}$ and \deff{column norm} $\norm{\cdot}{\col}$, given respectively as 
\begin{equation}\begin{aligned}
  \norm{\left(A_{E,x}\right)_x}{\row} \coloneqq \norm{ \sqrt{ \sum_x A_{E,x} A_{E,x}^\dagger } }{1} , \qquad \norm{(B_{E,x})_x}{\col} \coloneqq \norm{ \sqrt{ \sum_x B_{E,x}^\dagger B_{E,x} } }{1}.
\end{aligned}\end{equation}
The row--column norm then forms a symmetrised interpolation function between the two:
\begin{equation}\begin{aligned}
  \norm{\left(R_{E,x}\right)_x}{\rowcol} &= \inf_{R_{E,x} = A_{E,x} + B_{E,x}} \left[ \norm{(A_{E,x})_x}{\row} +  \norm{(B_{E,x})_x}{\col} \right].
\end{aligned}\end{equation}
These norms are precisely the ones that emerge in non-commutative Khintchine inequalities~\cite{lust-piquard_1991,pisier_1997,junge_2008,pisier_2009}.

As we now show, the Bures norm optimised over all states of one subsystem --- in other words, the norm underlying the conditional measured R\'enyi entropy of order 2 --- equals exactly this row--column norm.
Its natural smoothing variant, which we denoted $\hqr\pi_{k}(X|E)_\rho$, is then an interpolation functional between the space of trace-class operators and this row--column space.

\begin{boxed}
\begin{lemma}
For any bipartite Hermitian operator $R_{AE}$, it holds that
\begin{equation}\begin{aligned}
  \inf_{\sigma_E} \norm{R_{AE}}{\id_A \otimes \sigma_E} = \inf_{R_{AE} = \frac12(Z^{\vphantom{\dagger}}_{AE}+Z_{AE}^\dagger)} \Tr \sqrt{ \Tr_A\! \left( Z^{\vphantom{\dagger}}_{AE} Z_{AE}^\dagger\right) },
\end{aligned}\end{equation}
where the optimisation on the left-hand side is over all states $\sigma_E$, and on the right-hand side over all complex linear operators $Z_{AE}$.
Consequently, for a classical--quantum $R_{XE} = \sum_x \proj{x} \otimes R_{E,x}$ we have
\begin{equation}\begin{aligned}\label{eq:l1l2_cq}
   \exp\left( - \frac12 H_{2}^{\MM,\uparrow}(X|E)_R \right) &=  \inf_{\sigma_E} \norm{R_{XE}}{\id_X \otimes \sigma_E} \\&= \inf_{R_{E,x} = \frac12(Z^{\vphantom{\dagger}}_{E,x} + Z^\dagger_{E,x})} \Tr \sqrt{ \sum_x Z^{\vphantom{\dagger}}_{E,x} Z_{E,x}^\dagger }\\
  &= \norm{\left(R_{E,x}\right)_x}{\rowcol},
\end{aligned}\end{equation}
and for positive $\rho_{XE} = \sum_x \proj{x} \otimes \rho_{E,x} \geq 0$ it holds that
\begin{equation}\begin{aligned}
  \hqr\pi_{k}(X|E)_\rho &= \inf_{R_{XE}} \left[ \norm{\rho_{XE}-R_{XE}}{1} + \sqrt{k} \norm{\left(R_{E,x}\right)_x}{\rowcol} \right]\\
  &= \inf_{\rho^{\vphantom{\dagger}}_{E,x} = Q^{\vphantom{\dagger}}_{E,x} + \frac12 (Z^{\vphantom{\dagger}}_{E,x} + Z_{E,x}^\dagger)} \left( \sum_x \norm{Q_{E,x}}{1} + \sqrt{k} \Tr \sqrt{ \sum_x Z^{\vphantom{\dagger}}_{E,x} Z_{E,x}^\dagger } \right)\\
  &= K\left( \rho_{XE},\, \sqrt{k};\, L_1,\,  \rowcol) \right).
\end{aligned}\end{equation}
\end{lemma}
\end{boxed}
\begin{proof}
Let us consider for now general $R$ and $\sigma \geq 0$. 
We begin similarly as in Corollary~7 of~\cite{regula_2026}, establishing a dual expression
\begin{equation}\begin{aligned}
  \norm{R}{\sigma}^2 &\texteq{(i)} \sup \lset 2 \Tr H R - \Tr H^2 \sigma \bar H = H^\dagger \rset\\
  &\texteq{(ii)} \sup \lset 2 \Tr H R - \Tr C \sigma \bar H = H^\dagger,\; C \geq H^2 \rset\\
  &\texteq{(iii)} \sup \lset 2 \Tr H R - \Tr C \sigma \bar H = H^\dagger,\; \begin{pmatrix}C & H \\ H & \id \end{pmatrix} \geq 0 \rset\\
  &\texteq{(iv)} \inf \lset \Tr Q_{22} \bar Q = \begin{pmatrix} Q_{11} & Q_{12} \\ Q_{21} & Q_{22} \end{pmatrix} \geq 0,\; Q_{12} + Q_{21} = 2 R,\; Q_{11} = \sigma \rset.
\end{aligned}\end{equation}
Here, (i) is a variational form shown e.g.\ in the proof of~\cite[Lemma~4]{regula_2026}, (ii) follows since the positivity of $\sigma$ ensures that $\Tr C \sigma \geq \Tr H^2 \sigma$ for any $C \geq H^2$, (iii) is a standard Schur complement result~\cite[Thm.~1.3.3, Ex.~1.3.5]{bhatia_2007}, and (iv) is by strong Lagrange duality. 
For the second Schur complement, assume $\sigma>0$. By (i), $\norm{R}{\sigma+\delta\id}\uparrow\norm{R}{\sigma}$ as $\delta\downarrow0$; adding $\delta\id$ to $\sigma$ and renormalizing shows that the infimum below may be taken over strictly positive operators.
Using Schur complements again, the positivity of $Q$ is equivalent to the conditions that $Q_{22} \geq 0$, $Q_{21} = Q_{12}^\dagger$ and $Q_{22} \geq Q_{12}^\dagger Q_{11}^{-1} Q_{12}$, and hence 
\begin{equation}\begin{aligned}\label{eq:variational_quadratic}
  \norm{R}{\sigma}^2 &= \inf \lset \Tr Q_{22} \bar Q_{22} \geq Q_{12}^\dagger Q_{11}^{-1} Q_{12},\; Q^{\vphantom{\dagger}}_{12} + Q^\dagger_{12} = 2 R,\; Q_{11} = \sigma \rset\\
  &= \inf \lset \Tr Q_{12}^\dagger \sigma^{-1} Q_{12} \bar \Re(Q_{12}) = R \rset\\
  &= \inf \lset \Tr \sigma^{-1} Z Z^\dagger \bar \Re(Z) = R \rset,
\end{aligned}\end{equation}
where we denoted $Z = Q_{12}$ and $\Re(Z) = \frac12 (Z+Z^\dagger)$.

We now specialise to the bipartite case with $\sigma = \id_A \otimes \sigma_E$. Using that $(\id_A \otimes \sigma_E)^{-1} = \id_A \otimes \sigma_E^{-1}$, the reverse H\"older inequality (see e.g.~\cite[Lemma~3.3]{tomamichel_2016}) gives
\begin{equation}\begin{aligned}\label{eq:reverse_holder}
  \norm{R_{AE}}{\id_A \otimes \sigma_E}^2 &= \inf \lset \Tr \sigma_E^{-1} \Tr_A\!\left(Z^{\vphantom{\dagger}}_{AE} Z_{AE}^\dagger\right) \bar \Re(Z_{AE}) = R_{AE} \rset\\
  &\geq \inf \lset \norm{\sigma_E}{1}^{-1} \norm{\Tr_A\!\left(Z^{\vphantom{\dagger}}_{AE} Z_{AE}^\dagger\right)}{1/2} \bar \Re(Z_{AE}) = R_{AE} \rset
\end{aligned}\end{equation}
where $\norm{P}{1/2} = \big(\! \Tr \sqrt{P} \big)^2$ for $P \geq 0$. Since equality in~\eqref{eq:reverse_holder} can always be achieved by choosing
\begin{equation}\begin{aligned}
  \sigma_E = \frac{\sqrt{\Tr_A\!\left(Z^{\vphantom{\dagger}}_{AE} Z_{AE}^\dagger\right)}}{\Tr \sqrt{\Tr_A\!\left(Z^{\vphantom{\dagger}}_{AE} Z_{AE}^\dagger\right)}},
\end{aligned}\end{equation}
an optimisation over all states $\sigma_E$ leads to
\begin{equation}\begin{aligned}
  \inf_{\sigma_E} \norm{R_{AE}}{\id_A \otimes \sigma_E}^2 &= \inf \lset \norm{\Tr_A\!\left(Z^{\vphantom{\dagger}}_{AE} Z_{AE}^\dagger\right)}{1/2} \bar \Re(Z_{AE}) = R_{AE} \rset,
\end{aligned}\end{equation}
which is precisely the claimed expression. 

The expression for the CQ case in~\eqref{eq:l1l2_cq} is then immediate. Equality with the row--column norm
\begin{equation}\begin{aligned}
  \norm{\left(R_{E,x}\right)_x}{\rowcol} &= \inf_{R_{E,x} = A_{E,x} + B_{E,x}} \left( \norm{(A_{E,x})_x}{\row} +  \norm{(B_{E,x})_x}{\col} \right)\\
  &= \inf_{R_{E,x} = A_{E,x} + B_{E,x}} \left( \Tr \sqrt{ \sum_x A_{E,x} A_{E,x}^\dagger } + \Tr \sqrt{ \sum_x B_{E,x}^\dagger B_{E,x} } \right)
\end{aligned}\end{equation}
follows because for Hermitian $R_{XE}$ it suffices to consider decompositions with $B_{E,x} = A_{E,x}^\dagger$; this can be seen by noting that given any decomposition $R_{E,x} = A_{E,x} + B_{E,x} = A^\dagger_{E,x} + B^\dagger_{E,x}$ we can define $Z_{E,x} = A_{E,x} + B^\dagger_{E,x}$ so that $R_{E,x} = \frac12 \left( Z_{E,x} + Z_{E,x}^\dagger \right)$ and
\begin{equation}\begin{aligned}
 \norm{(Z_{E,x})_x}{\row}  &= \frac12 \left( \norm{(Z_{E,x})_x}{\row} + \norm{(Z^\dagger_{E,x})_x}{\col} \right)\\ 
 &\leq \frac12 \left( \norm{(A_{E,x})_x}{\row} + \norm{(B^\dagger_{E,x})_x}{\row} + \norm{(A^\dagger_{E,x})_x}{\col} + \norm{(B_{E,x})_x}{\col} \right)\\
 &= \norm{(A_{E,x})_x}{\row} + \norm{(B_{E,x})_x}{\col}
\end{aligned}\end{equation}
which means that the decomposition $R_{E,x} = \frac12 \left( Z_{E,x} + Z_{E,x}^\dagger \right)$ is at least as good as $R_{E,x} = A_{E,x} + B_{E,x}$. 

Finally, for a positive CQ operator $\rho_{XE}$, the result follows by combining the above with the variational form given in Lemma~\ref{lem:lhl}. 
\end{proof}

\begin{remark}
From Eq.~\eqref{eq:variational_quadratic} we can also notice a rather intriguing variational form:
\begin{equation}\begin{aligned}
  D_2^{\MM}(\rho\|\sigma) = \inf \lset \ol{D}_2\big(\sqrt{ZZ^{\smash\dagger}} \big\| \sigma\big) \bar \rho = \Re(Z) \rset,
\end{aligned}\end{equation}
where we recall the Petz--R\'enyi collision divergence $\ol{D}_2(\omega\|\sigma) = \log \Tr \omega^2 \sigma^{-1}$.
\end{remark}


\subsection{Remarks on 2-universal hashing and balanced hash functions}

Let us briefly return to the simple case where the side information system is taken to be trivial, i.e.\ the source is described by a probability distribution $p_X$.
Recalling the intrinsic randomness exponent discussed in Section~\ref{sec:warmup}, we know that an optimisation over all hash functions yields
\begin{equation}\begin{aligned}\label{eq:intrinsic_exponent_allhash}
   \lim_{n\to\infty} - \frac1n \log \left[ \Delta_{\min}(p^{\otimes n}_{X^n}, \exp(Rn)) \right] =  \sup_{\alpha \in (1,2]} \,(\alpha-1) \Big( H_{\alpha}(X)_p - R \Big)
\end{aligned}\end{equation}
for sufficiently large rates $R > R_{\rm crit}$.  
On the other hand, Theorem~\ref{thm:reallytight} tells us that a restriction to 4*-universal hashing gives
\begin{equation}\begin{aligned}\label{eq:intrinsic_exponent_4univ}
   \lim_{n\to\infty} - \frac1n \log \left[ \Delta_{\min,\univstars{4}}(p^{\otimes n}_{X^n}, \exp(Rn)) \right] =  \sup_{\alpha \in (1,2]} \frac{\alpha-1}{\alpha} \Big( H_{\alpha}(X)_p - R \Big)
\end{aligned}\end{equation}
for all $R > 0$.
The discrepancy explicitly shows that the exponents in the two regimes differ --- a property previously observed in a weaker form by~\cite{hayashi_2013}. 

One may then wonder whether the assumption of 4*-universal hash functions is necessary for the exponent to take the form as in~\eqref{eq:intrinsic_exponent_4univ}. Although we are not able to resolve the question of the weakest possible assumption for the converse of Theorem~\ref{thm:reallytight} to apply, we will now show that the assumption of 2-universality is not sufficient: an exponent larger than~\eqref{eq:intrinsic_exponent_4univ} can be achieved under 2-universal hash functions for some sources and rates. 
This is due to the fact that balanced hash functions, namely ones with approximately equally-sized bins, can always be chosen to be 2-universal, while at the same time performing better for certain sources. The simplest example is a flat (uniform) source, from which even zero-error randomness extraction is possible (see for example~\cite[Example~2]{li_2023}). 
Indeed, this can be studied more generally by extending the simple min-entropy--based bound from Lemma~\ref{lem:minent} to work in a randomised way, as long as the source can be ensured to be sufficiently uniform --- which, importantly, is the case for i.i.d.\ sources.

\begin{boxed}[white]
\begin{lemma}[Balanced hashing of flat sources]\label{lem:balanced}
Assume that the source distribution $p_X$ consists of $K$ uniform sub-distributions, i.e.\ $p_X = \sum_{k=1}^K p_{S_k}$ where $p_{S_k}(x) = \lambda_k$ for all $x \in S_k$ with disjoint sets $S_k$ such that $\bigcup_{k=1}^K S_k = \X$.

\smallskip
Then there exists a 2-universal family of hash functions that achieves the one-shot bound
\begin{equation}\begin{aligned}
  \EE_f \frac12 \norm{ p^f_Z - \frac{\id}{|\Z|}}{1} &\leq \inf_{\ve \in [0,1)}\left[ \ve + K \, |\Z| \exp(-H_{\min}^\ve(X)_p) \right]\\
  &\leq K \, \pi_{|\Z|}(p_X \| \id_X).
\end{aligned}\end{equation}
Consequently, for any i.i.d.\ source $p_{X^n}^{\otimes n}$, 2-universal hashing achieves the error exponent
\begin{equation}\begin{aligned}\label{eq:balaned_exponent}
   \liminf_{n\to\infty} - \frac1n \log \left[ \Delta_{\min,\univ{2}}(p^{\otimes n}_{X^n}, \exp(nR)) \right] \geq \sup_{\alpha > 1} \frac{\alpha-1}{\alpha} \Big( H_{\alpha}(X)_p - R \Big).
\end{aligned}\end{equation}
\end{lemma}
\end{boxed}
\begin{proof}
Consider first a simple hashing procedure $h \colon \X \to \Z$ where we aim to make each of the $|\Z|$ bins balanced, that is, to assign them an approximately equal number of elements; this can be done by choosing at most $\left\lceil |S_k|/|\Z| \right\rceil$ elements from each of the uniform subsets $S_k$. Applying this to $p_X$ directly would then give a bound on the distance from uniform as
\begin{equation}\begin{aligned}\label{eq:balanced_Hmin}
  \frac12 \norm{ p^h_Z - \frac{\id}{|\Z|}}{1} &\leq \sum_z \left[ \sum_k \left( \frac{|S_k|}{|\Z|} + 1 \right) \lambda_k - \frac{1}{|\Z|}\right]_+\\
  &= \sum_z \sum_k \lambda_k\\
  &\leq \sum_z K \max_k \lambda_k\\
  &= |\Z| \, K \, \exp(-H_{\min}(X)_p).
\end{aligned}\end{equation}
For a $p_X$ satisfying the assumptions of the lemma, an optimal $p'_X$ in the definition of $H_{\min}^\ve$ can be chosen as $p'_X = p_X \wedge \lambda \id$ which also has the same uniform structure, so applying the argument to $p'_X$ instead gives the smooth entropy bound as in Lemma~\ref{lem:minent}; the hypothesis testing bound then follows as in Corollary~\ref{cor:hyp_test_bound}.

It thus remains to argue that this can be done in a 2-universal manner. For this, let $\pi_k$ be a uniformly random permutation of the elements of each $S_k$, let $\sigma_k$ be an independent uniformly random permutation of $\Z$, and define the randomised choice of the hash function as $f(x) \coloneqq \sigma_k ( h ( \pi_k (x) ) )$ for each $x \in S_k$. Clearly, none of the bounds in~\eqref{eq:balanced_Hmin} are affected by this, since the elements of $p$ on each $S_k$ are all equal by assumption, and the specific choice of a specific bin $z \in \Z$ is immaterial. But this randomisation is indeed enough to give 2-universality: for $x \neq x'$ with $x \in S_k  \not= S_{k'} \ni x'$, the uniformly random choice ensures that $\Pr[f(x) = f(x')] = \frac{1}{|\Z|}$, while for $x, x' \in S_k$ we have
\begin{equation}\begin{aligned}
  \Pr[f(x) = f(x')] &= \sum_z \frac{ |h^{-1}(z) \cap S_k| \left( |h^{-1}(z) \cap S_k| - 1\right)}{|S_k|(|S_k|-1)}\\
  &\leq \sum_z \frac{|h^{-1}(z) \cap S_k| \left(  \left\lceil \frac{|S_k|}{|\Z|} \right\rceil - 1\right)}{|S_k|(|S_k|-1)}\\
  &=\frac{|S_k| \left(  \left\lceil \frac{|S_k|}{|\Z|} \right\rceil - 1\right)}{|S_k|(|S_k|-1)}\\
  &=\frac{\left\lceil \frac{|S_k|}{|\Z|} \right\rceil - 1}{|S_k|-1}\\
  &\leq \frac{1}{|\Z|}.
\end{aligned}\end{equation}

If the source is i.i.d., then the required uniformity is guaranteed on its type classes. Specifically, we can write $p_{X^n}^{\otimes n} = \sum_{k} \Pr_{p^{\otimes n}} [ T_k ] \, u_k$ where $T_k$ denotes a type and $u_k$ a distribution uniform on the corresponding type class (see e.g.~\cite{csiszar_2011}). A standard type-counting argument gives $K \leq (n+1)^{|\X|}$, which does not affect the asymptotic error exponent: we get
\begin{equation}\begin{aligned}
  \EE_{f_n} \frac12 \norm{ \left(p^{\otimes n}_{X^n}\right)^{f_n} - \frac{\id}{|\Z^n|}}{1} &\leq (n+1)^{|\X|} \,|\Z^n|^{\frac{\alpha-1}{\alpha}} \exp\!\left(- n \frac{\alpha-1}{\alpha} H_\alpha(X)_p\right)
\end{aligned}\end{equation}
for all $\alpha > 1$, from which~\eqref{eq:balaned_exponent} follows by setting $|\Z^n| = \exp(nR)$.
\end{proof}

The bound of Lemma~\ref{lem:balanced} can exceed~\eqref{eq:intrinsic_exponent_4univ} for small rates $R$, which is when the optimal $\alpha$ in the optimisation on the right-hand side of~\eqref{eq:balaned_exponent} may satisfy $\alpha > 2$.

However, a limitation of Lemma~\ref{lem:balanced} is that it appears unclear how to extend it to the case of side information, even in the classical case, as the `balancing' of bins would need to account for the whole side information system. It is thus not obvious to us whether an exponent of the form in~\eqref{eq:balaned_exponent} can be achieved with universal hash functions for more general cases, and in particular for quantum side information. 
Indeed, a relevant prior work here is~\cite{shen_2023-1} that studied privacy amplification and the closely related task of soft covering for constant-type sources; there, only the exponent range $\alpha \in (1,2]$ is achieved precisely due to the presence of side information. 
This is one reason why the analysis of universal arguments that apply to general sources, such as the leftover hash lemma, is key to the study of randomness extraction against quantum side information. The advantage of the converse bound of Theorem~\ref{thm:reallytight} is precisely that it makes no assumptions about the source and could accurately describe the behaviour of hashing for general quantum states, although the precise conditions for applicability may require more structural assumptions about the hash functions in consideration.


\section{Strong converse exponent under 2-universal hashing}
\label{sec:strong-converse}

In the strong converse regime we are concerned with the decay of
\begin{equation}\begin{aligned}
  1 - \EE_f \norm{\rho_{ZE}^f  - \frac{\id_Z}{|\Z|} \otimes \rho_E}{+} = \EE_f\, \Tr \left[ \rho_{ZE}^f  \wedge \frac{\id_Z}{|\Z|} \otimes \rho_E \right].
\end{aligned}\end{equation}
One way to get an achievability bound here is from the leftover hash lemma (Lemma~\ref{lem:lhl}): this can be written as
\begin{equation}\begin{aligned}
  \EE_f \,\Tr \left[ \rho_{ZE}^f  \wedge \frac{\id_Z}{|\Z|} \otimes \rho_E \right] \geq \sup_{\sigma_E} \,\hq\beta_{|\Z|/2}  \!\left( \id_X \otimes \sigma_E \middle\|  \rho_{XE} \right)
\end{aligned}\end{equation}
for all 2-universal hash functions, 
where we denoted $\hq\beta_{\ve}(B\|A) = \Tr A - \hqr\pi_\ve(A\|B)$. 
Through connections with exponents of conventional hypothesis testing~\cite{audenaert_2008} it can then be shown that
\begin{equation}\begin{aligned}
    \lim_{n\to\infty} -\frac1n \log \hq\beta_{\exp(-n R)} (\sigma^{\otimes n} \| \rho^{\otimes n})  = \sup_{\alpha \in (0,1)} \frac{\alpha-1}{\alpha} \left( R -  \ol{D}_{\alpha}(\rho\|\sigma) \right),
\end{aligned}\end{equation}
which means that the best bound on the strong converse exponent for privacy amplification that can be obtained from a direct application of the leftover hash lemma is exactly as given already in~\cite[Proposition~26]{regula_2026}, i.e.\ $\sup_{\alpha \in (0,1)} \frac{\alpha-1}{\alpha} \left( \ol{H}^{\uparrow}_{\alpha}(X|E)_\rho - R \right)$. This is however not optimal.

We now give an improved achievability result by combining the techniques underlying the leftover hash lemma of~\cite{regula_2026} with half-quadratic hypothesis testing. We show also that this is approximately optimal, yielding a tight asymptotic result.

\begin{tboxed}[filled]{Tight one-shot bounds on strong converse error}
\begin{theorem}\label{thm:strongconverse}
For any 2\*-universal family of hash functions, it holds that
\begin{equation}\begin{aligned}
\EE_f \,\Tr \!\left[ \rho_{ZE}^f  \wedge \frac{\id_Z}{|\Z|} \otimes \rho_E \right] \geq \frac12 \Tr \!\left[ \rho_{XE}  \wedge  \frac{\id_X}{|\Z|} \otimes \rho_E \right],
\end{aligned}\end{equation}
while for 2-universal hash families we have
\begin{equation}\begin{aligned}\label{eq:strongconverse_achievability}
\EE_f \,\Tr \!\left[ \rho_{ZE}^f  \wedge \frac{\id_Z}{|\Z|} \otimes \rho_E \right] \geq \frac14 \Tr \!\left[ \rho_{XE}  \wedge  \frac{\id_X}{|\Z|} \otimes \rho_E \right].
\end{aligned}\end{equation}
These bounds are optimal among 2-universal hashes up to a constant factor: for any 2-universal family,
\begin{equation}\begin{aligned}\label{eq:strongconverse_converse}
\EE_f \,\Tr \!\left[ \rho_{ZE}^f  \wedge \frac{\id_Z}{|\Z|} \otimes \rho_E \right] \leq 4 \Tr \!\left[ \rho_{XE}  \wedge  \frac{\id_X}{|\Z|} \otimes \rho_E \right].
\end{aligned}\end{equation}

This shows that the strong converse exponent under 2-universal hash functions is exactly
\begin{equation}\begin{aligned}\label{eq:sc-exponent-thm}
 \lim_{n\to\infty} - \frac1n \log \left[ 1 - \Delta_{\min, \univ{2}}\!\left(\rho_{XE}^{\otimes n},\, \exp(nR) \right)\right] = \sup_{\alpha \in (0,1)} (\alpha-1) \left( \ol{H}^{\downarrow}_{\alpha}(X|E)_\rho - R \right),
\end{aligned}\end{equation}
and it is achieved by any 2-universal hash family.
\end{theorem}
\end{tboxed}

\begin{remark}
The asymptotics in~\eqref{eq:sc-exponent-thm} are most easily obtained by using the Nussbaum--Szkoła~\cite{nussbaum_2009} distributions corresponding to $\rho_{XE}$ and $\frac{\id_X}{|\X|} \otimes \rho_E$. Specifically, writing the states in their spectral decompositions as $\rho_{XE} = \sum_i a_i \proj{\alpha_i}$ and $\frac{\id_X}{|\X|} \otimes \rho_E = \sum_j b_j \proj{\beta_j}$, the Nussbaum--Szkoła distributions $p$ and $q$ are given by $p(i,j) = a_i \left|\braket{\alpha_i | \beta_j}\right|^2$ and $q(i,j) = b_j \left|\braket{\alpha_i | \beta_j}\right|^2$. 
Combining our result with the one-shot bounds of~\cite{cheng_2026}, we have that
\begin{equation}\begin{aligned}
  8 \Tr \!\left[ p \wedge \frac{|\X|}{|\Z|} q \right] \geq \EE_f \,\Tr \!\left[ \rho_{ZE}^f  \wedge \frac{\id_Z}{|\Z|} \otimes \rho_E \right] \geq \frac18 \Tr \!\left[ p \wedge \frac{|\X|}{|\Z|} q \right].
\end{aligned}\end{equation}
for all 2-universal hash functions. 
The exponential behaviour of the classical quantity $\Tr[p \wedge \frac{|\X|}{|\Z|} q]$ then follows from an application of standard large-deviation bounds~\cite{chernoff_1952}, giving an expression in terms of $D_\alpha(p\|q)$. Crucially, $D_\alpha (p \| q) = \ol{D}_\alpha\!\left(\rho_{XE} \middle\| \frac{\id_X}{|\X|} \otimes \rho_E\right)$~\cite{nussbaum_2009}, which yields~\eqref{eq:sc-exponent-thm}. 

Moreover, the tightness of our one-shot bounds can be used to obtain precise estimates on the speed of the convergence of the exponent to its asymptotic form. Specifically, 
with $|\Z^n| = \exp(nR + O(1))$ and $
H(X|E)_\rho < R <
\left.\frac{\mathrm d}{\mathrm d\alpha}
\bigl[(\alpha-1)\ol{H}^{\downarrow}_\alpha(X|E)_\rho\bigr]
\right|_{\alpha=0^+}$,
our one-shot bounds,
\cite[Eq.~(59), Theorem~5.2]{cheng_2026}, and the classical strong large deviation principle imply a third-order asymptotic expansion
\begin{equation}
\begin{aligned}
-\log\!\left[
1-\Delta_{\min,\univ{2}}\!\left(\rho_{XE}^{\otimes n},|\Z^n|\right)
\right]
&= n\sup_{\alpha\in(0,1)}(\alpha-1)
   \bigl(\ol{H}^{\downarrow}_\alpha(X|E)_\rho-R\bigr)
+\frac12\log n+O(1).
\end{aligned}
\end{equation}
\end{remark}

\begin{remark}
The converse bound of~\eqref{eq:strongconverse_converse} together with Audenaert's inequality~\cite{audenaert_2007,audenaert_2008} gives
\begin{equation}\begin{aligned}
  \EE_f \,\Tr \!\left[ \rho_{ZE}^f  \wedge \frac{\id_Z}{|\Z|} \otimes \rho_E \right] \leq 4\, |\Z|^{\alpha-1} \exp(-(\alpha-1) \,\ol{H}_{\alpha}^\downarrow(X|E)_\rho)
\end{aligned}\end{equation}
for all $\alpha \in (0,1)$, which exactly recovers the bound of~\cite{shen_2022} and shows that it holds also under 2-universal hashes without requiring pairwise independence.
\end{remark}

We will split the proof into two parts.

\subsection{Achievability through half-quadratic error}

Before we begin, consider the total (average) error in hypothesis testing:
\begin{equation}\begin{aligned}\label{eq:wedge}
  \err(A\|B) &\coloneqq \Tr [A \wedge B] \\
  &\hphantom{:}= \inf \lset \Tr (\id - M) A + \Tr M B \bar 0 \leq M \leq \id \rset\\
  &\hphantom{:}= \sup \lset \Tr R \bar R \leq A,\; R \leq B \rset.
\end{aligned}\end{equation}
We define and characterise its half-quadratic variant.
\begin{tboxed}[white]{Half-quadratic error}
\begin{lemma}\label{lem:strongconv_lemma}
For any positive semidefinite $A$ and $B$, define the average \deff{half-quadratic error} as
\begin{equation}\begin{aligned}\label{eq:halfquad_primal}
  \hqr\err(A\|B) &\coloneqq \inf \lset \Tr (\id - M) A + \Tr M^2 B \bar 0 \leq M \leq \id \rset.
\end{aligned}\end{equation}
Then
\begin{equation}\begin{aligned}\label{eq:halfquad_dual}
  \hqr\err(A\|B) = \sup \lset \Tr R - \frac14 \norm{R}{B}^2 \bar R \leq A \rset,
\end{aligned}\end{equation}
where we recall that $\norm{R}{B}^2 = \Tr R \J_{B}^{-1}(R)$.\medskip

Furthermore, there exist an optimal solution $R$ for~\eqref{eq:halfquad_dual} and an optimal measurement $M$ for~\eqref{eq:halfquad_primal} that satisfy
\begin{enumerate}[(i),nosep]
\item stationarity:\; $R = 2 \J_B(M)$,
\item complementary slackness: \;$\Tr(\id - M) A = \Tr(\id - M) R$.
\end{enumerate}
\medskip

The half-quadratic error can be related to standard hypothesis testing errors as
\begin{equation}\begin{aligned}
  \err(A\|B) \geq \;&\hqr\err(A\|B) \geq \frac34 \err(A\|B),\\
  2\err(p\|q) \geq \;&\hqr\err(A\|B) \geq \frac12 \err(p\|q)
\end{aligned}\end{equation}
with $p$, $q$ the Nussbaum--Szkoła distributions of $A$, $B$.
\end{lemma}
\end{tboxed}

\begin{proof}
Write
\begin{equation}\begin{aligned}
  \hqr\err(A\|B) &= \inf \lset \Tr (\id - M) A + \Tr \J_B(M) M \bar 0 \leq M \leq \id \rset,
\end{aligned}\end{equation}
recalling that $\J_{B}(M) = \frac12 (BM + MB)$. 
We first simplify our life a bit by noticing that the constraint $M \geq 0$ is superfluous and can be omitted without loss of generality. 
Then the Lagrange dual function evaluated at the Lagrange multiplier $P \geq 0$ is
\begin{equation}\begin{aligned}\label{eq:lagrangian}
 \inf_{M=M^\dagger} L(M;\,P) &= \inf_{M=M^\dagger} \Tr (\id - M) A + \Tr \J_B(M) M - \Tr (\id - M) P\\
 &\texteq{(i)} \inf_{Z=Z^\dagger} \Tr \!\left(\id - \J_{B}^{-1}(Z)\right) \!A + \Tr Z \J_{B}^{-1}(Z) - \Tr \!\left(\id - \J_{B}^{-1}(Z)\right) \!P\\
 &= \inf_{Z=Z^\dagger} \< P - A, Z \>_B + \norm{Z}{B}^2 + \Tr (A-P)\\
 &\texteq{(ii)} \inf_{Z=Z^\dagger} \norm{Z - \frac{A-P}{2}}{B}^2 - \norm{\frac{A-P}{2}}{B}^2 + \Tr (A-P)\\
 &= - \norm{\frac{A-P}{2}}{B}^2 + \Tr (A-P),
\end{aligned}\end{equation}
where in (i) we defined $Z = \J_{B}(M)$ and in (ii) we completed the square. 
Defining $R = A - P$, the dual form (Eq.~\eqref{eq:halfquad_dual}) follows by strong duality, seen e.g.\ by Slater's theorem as the primal is clearly strongly feasible. The existence of an optimal $R$ is guaranteed from the strong feasibility of the primal, and the infimum in the primal itself is easily seen to be achieved since the set $\lset M \bar 0 \leq M \leq \id \rset$ is compact. From~\eqref{eq:lagrangian} we also see that an optimiser must satisfy $Z = \frac{A-P}{2}$, and so it suffices to optimise over $R$ of the form $R = 2 Z = 2 \J_B(M)$ --- this is precisely the Karush--Kuhn--Tucker stationarity condition for this problem. For an optimal pair of $M$ and $R = 2 \J_B(M)$ we then notice that
\begin{equation}\begin{aligned}
 \norm{R}{B}^2 = 4 \Tr \J_B(M) M = 4 \Tr M^2 B = 2 \Tr M R,
\end{aligned}\end{equation}
from which it follows by the fact that $R \leq A$ that
\begin{equation}\begin{aligned}
  \Tr (\id - M) A + \Tr M^2 B &\geq \Tr (\id - M) R + \Tr M^2 B\\
  &= \Tr R - \Tr MR + \Tr M^2 B\\
  &= \Tr R - \frac14 \norm{R}{B}^2\\
  &= \Tr (\id - M) A + \Tr M^2 B,
\end{aligned}\end{equation}
implying the complementary slackness condition $\Tr (\id - M) R = \Tr (\id - M) A$.

Now, the inequality $\err(A\|B) \geq \hqr\err(A\|B)$ is immediate, since $M^2 \leq M$ for any feasible measurement $0 \leq M \leq \id$. 
For the opposite inequality, take any measurement operator $M$ and define $M' \coloneqq M^2 (M^2 - M + \id)^{-1}$. Since $t^2 - t + 1 = \left(t - \tfrac12\right){\!}^2 + \frac34\geq \frac34$, the commutativity of the involved operators ensures that $0 \leq M' \leq \frac43 M^2$ and $\id - M' = (\id - M)(M^2 - M + \id)^{-1} \leq \frac43 (\id - M)$.  Furthermore, $M \leq \id$ gives $(M^2 - M + \id) \geq M^{2}$ and hence that $M' \leq \id$.
Thus
\begin{equation}\begin{aligned}
  \frac43\, \hqr\err(A\|B) &= \inf \lset \Tr \frac43 (\id - M) A + \Tr \frac43 M^2 B \bar 0 \leq M \leq \id \rset\\
  &\geq \inf \lset \Tr (\id - M') A + \Tr M' B \bar 0 \leq M \leq \id \rset\\
  &\geq \inf \lset \Tr (\id - M') A + \Tr M' B \bar 0 \leq M' \leq \id \rset\\
  &= \err(A\|B).
\end{aligned}\end{equation}

Finally, \cite[Theorem~4.2]{cheng_2026} showed that $\err(A\|B) \leq 2\err(p\|q)$, and we also have
\begin{equation}\begin{aligned}
  \hqr\err(A\|B) &= \inf \lset \Tr (\id - M) A + \Tr M^2 B \bar 0 \leq M \leq \id \rset\\
  &\geq \inf \lset \Tr (\id - M)^2 A + \Tr M^2 B \bar 0 \leq M \leq \id \rset\\
  &\geq \inf\, \Tr \!\big[ (\id - X) A (\id - X)^\dagger \big] + \Tr \!\big[ X^\dagger B X \big] \\
  &\geq \frac12 \err(p\|q)
\end{aligned}\end{equation}
with the last inequality shown in~\cite[Appendix~B]{cheng_2026}. 
\end{proof}

We are now ready to prove the achievability statement, which we state as a separate lemma for clarity.

\begin{boxed}
\begin{lemma}[{Achievability of Theorem~\ref{thm:strongconverse}}] \label{lemma:strongconverse_achievability}
For any 2\*-universal family of hash functions, it holds that
\begin{equation}\begin{aligned}
\EE_f \,\Tr \!\left[ \rho_{ZE}^f  \wedge \frac{\id_Z}{|\Z|} \otimes \rho_E \right] \geq \frac12 \Tr \!\left[ \rho_{XE}  \wedge  \frac{\id_X}{|\Z|} \otimes \rho_E \right],
\end{aligned}\end{equation}
while for 2-universal hash functions we have
\begin{equation}\begin{aligned}
\EE_f \,\Tr \!\left[ \rho_{ZE}^f  \wedge \frac{\id_Z}{|\Z|} \otimes \rho_E \right] \geq \frac14 \Tr \!\left[ \rho_{XE}  \wedge  \frac{\id_X}{|\Z|} \otimes \rho_E \right].
\end{aligned}\end{equation}
\end{lemma}
\end{boxed}

\begin{proof}
By Lemma~\ref{lem:strongconv_lemma}, we have
\begin{equation}\begin{aligned}\label{eq:strongconv_step1}
  \EE_f \left( 1 - \frac12\norm{\rho_{ZE}^f  - \frac{\id_Z}{|\Z|} \otimes \rho_E}{1} \right) &= \EE_f \, \err\left(\rho_{ZE}^f  \middle\| \frac{\id_Z}{|\Z|} \otimes \rho_E \right) \\
  &\geq \EE_f \,\hqr\err\left(\rho_{ZE}^f  \middle\| \frac{\id_Z}{|\Z|} \otimes \rho_E \right)\\
  &= \EE_f \sup \lset \Tr R_{ZE} - \frac{|\Z|}{4} \norm{R_{ZE}}{\id_Z \otimes \rho_E}^2 \bar R_{ZE} \leq \rho_{ZE}^f \rset.
\end{aligned}\end{equation}
Consider now that, for any classical-quantum $R_{XE} \leq \rho_{XE}$, the corresponding operator $R_{ZE}^f \coloneqq \sum_z \proj{z} \otimes \sum_{x : f(x) = z} R_{E,x}$ clearly satisfies $R_{ZE}^f \leq \rho_{ZE}^f$ as the hashing map $f$ is a valid quantum channel; $R_{ZE}^f$ is then feasible for~\eqref{eq:strongconv_step1}. Crucially, the proof of the leftover hash lemma in~\cite[Lemma~12]{regula_2026} (Eq.~(92)) tells us that any 2*-universal hash family satisfies
\begin{equation}\begin{aligned}\label{eq:lhl}
  \EE_f \norm{R_{ZE}^f}{\id_Z \otimes \rho_E}^2 &= \left(1 - \frac{1}{|\Z|}\right) \norm{R_{XE}}{\id_X \otimes \rho_E}^2 + \frac{1}{|\Z|} \norm{R_E}{\rho_E}^2.
\end{aligned}\end{equation}
In order to bound the rightmost term here, we will restrict our attention to $R_{XE}$ which are optimisers of $\hqr\err(\rho_{XE} \| k \id_X \otimes \rho_E)$ for some $k > 0$. As we showed in Lemma~\ref{lem:strongconv_lemma}, any such optimiser takes the form $R_{XE} = 
2 \J_{k \id_X \otimes \rho_E}(M_{XE})$ for some optimal measurement operator $M_{XE}$. The key reason why this is needed is because, even though $R_{XE} \not\geq 0$ in general, it does hold that $\J_{\id_X \otimes \rho_E}^{-1}(R_{XE}) = k \J_{k \id_X \otimes \rho_E}^{-1}(R_{XE}) = 2k\, M_{XE} \geq 0$. This allows us to deduce that
\begin{equation}\begin{aligned}\label{eq:uhhhhh}
  \Tr R_{XE} &= \Tr R_{XE} \id_{XE} \\
  &= \Tr R_{XE}\, \J_{\id_X \otimes \rho_{E}}^{-1}\!\left( \id_X \otimes \rho_E\right)\\
  &= \Tr \J_{\id_X \otimes \rho_{E}}^{-1}\!\left(R_{XE}\right) \id_X \otimes \rho_E\\
  &\geq \Tr \J_{\id_X \otimes \rho_{E}}^{-1}\!\left(R_{XE}\right) \id_X \otimes R_E\\
  &= \Tr R_{XE} \big[\id_X \otimes \J^{-1}_{\rho_E} (R_E)\big]\\
  &= \norm{R_E}{\rho_E}^2,
\end{aligned}\end{equation}
where we used that $R_{XE} \leq \rho_{XE}$ and that $\J_{\id_X \otimes \rho_E} = \idc_X \otimes \J_{\rho_E}$.

Plugging~\eqref{eq:lhl} into~\eqref{eq:strongconv_step1} and using~\eqref{eq:uhhhhh} then gives
\begin{equation}\begin{aligned}\label{eq:almostdone}
  \EE_f \, \err\left(\rho_{ZE}^f  \middle\| \frac{\id_Z}{|\Z|} \otimes \rho_E \right) &\geq \Tr R^f_{ZE} - \frac{|\Z|}{4} \EE_f \norm{R_{ZE}^f}{\id_Z \otimes \rho_E}^2\\
  &= \Tr R_{XE} - \frac{|\Z|-1}{4} \norm{R_{XE}}{\id_X \otimes \rho_E}^2 - \frac14 \norm{R_E}{\rho_E}^2\\
  &\geq  \Tr R_{XE} - \frac{|\Z|}{4} \norm{R_{XE}}{\id_X \otimes \rho_E}^2 - \frac14 \Tr R_{XE}\\
  &= \frac34 \Tr R_{XE} - \frac{|\Z|}{4} \norm{R_{XE}}{\id_X \otimes \rho_E}^2.
\end{aligned}\end{equation}
From here, it is already possible to deduce that 
\begin{equation}\begin{aligned}
  \EE_f \, \err\left(\rho_{ZE}^f  \middle\| \frac{\id_Z}{|\Z|} \otimes \rho_E \right) &\geq \frac{9}{16}\, \hqr\err \left( \rho_{XE}  \middle\| \frac{\id_X}{|\Z|} \otimes \rho_E \right) \geq \frac{27}{64} \,\err \left( \rho_{XE}  \middle\| \frac{\id_X}{|\Z|} \otimes \rho_E \right)
\end{aligned}\end{equation}
through a use of Lemma~\ref{lem:strongconv_lemma} and some manipulation. However, in this case we can do a bit better than that.
Namely, let us assume that $R_{XE}$ is specifically an optimiser of $\hqr\err(\rho_{XE} \| \frac{\id_X}{|\Z|} \otimes \rho_E)$. The complementary slackness condition of Lemma~\ref{lem:strongconv_lemma} tells us that $\Tr (\id - M_{XE}) \rho_{XE} = \Tr (\id - M_{XE}) R_{XE}$ holds for an optimal operator $R_{XE}$ and an optimal measurement $M_{XE}$ such that $2 \J_{\id_X/|\Z|  \otimes \rho_E}\!(M_{XE}) = R_{XE}$. This choice of measurement then gives
\begin{equation}\begin{aligned}\label{eq:reallydone}
  \err\left( \rho_{XE} \middle\| \frac{\id_X}{|\Z|} \otimes \rho_E \right) &\leq \Tr (\id - M_{XE}) R_{XE} + \Tr M_{XE} \frac{\id_X}{|\Z|}\otimes \rho_E \\
  &= \Tr R_{XE} - \frac{1}{2} \Tr \!\left[ \J_{\frac{\id_X}{|\Z|}\otimes \rho_E}^{-1}\!(R_{XE})\, R_{XE} \right] + \frac{1}{2} \Tr \!\left[ \J_{\frac{\id_X}{|\Z|} \otimes \rho_E}^{-1}\!(R_{XE})\, \frac{\id_X}{|\Z|}\otimes \rho_E \right]\\
  &= \Tr R_{XE} - \frac12 \norm{R_{XE}}{\frac{\id_X}{|\Z|} \otimes \rho_E}^2 + \frac{1}{2} \Tr \!\left[ R_{XE} \,\J_{\frac{\id_X}{|\Z|} \otimes \rho_E}^{-1}\!\left(\frac{\id_X}{|\Z|}\otimes \rho_E\right) \right]\\
  &=  \Tr R_{XE} - \frac{|\Z|}{2} \norm{R_{XE}}{\id_X \otimes \rho_E}^2 + \frac12 \Tr R_{XE}\\
  &= 2 \left( \frac34 \Tr R_{XE} - \frac{|\Z|}{4} \norm{R_{XE}}{\id_X \otimes \rho_E}^2 \right)\\
  &\leq 2 \, \EE_f \, \err\left(\rho_{ZE}^f  \middle\| \frac{\id_Z}{|\Z|} \otimes \rho_E \right)
\end{aligned}\end{equation}
by~\eqref{eq:almostdone}, which is precisely the desired bound.

The case of 2-universal hash functions proceeds analogously, with the only difference being that now we have to use the weaker inequality~\cite[Proposition~7]{tan_2026}
\begin{equation}\begin{aligned}\label{eq:lhl_2univ}
  \EE_f \norm{R_{ZE}^f}{\id_Z \otimes \rho_E}^2 &\leq 2 \left(1 - \frac{1}{|\Z|}\right) \norm{R_{XE}}{\id_X \otimes \rho_E}^2 + \frac{1}{|\Z|} \norm{R_E}{\rho_E}^2,
\end{aligned}\end{equation}
which yields
\begin{equation}\begin{aligned}
  \EE_f \, \err\left(\rho_{ZE}^f  \middle\| \frac{\id_Z}{|\Z|} \otimes \rho_E \right) &\geq   \frac34 \Tr R_{XE} - \frac{|\Z|}{2} \norm{R_{XE}}{\id_X \otimes \rho_E}^2.
\end{aligned}\end{equation}
Choosing $R_{XE}$ as an optimiser of $\hqr\err(\rho_{XE} \| \frac14 \frac{\id_X}{|\Z|} \otimes \rho_E)$ and repeating the steps of~\eqref{eq:reallydone} gives the claimed inequality.
\end{proof}

\subsection{Converse through quadratic error}

For the converse statement, we will employ the \deff{fully quadratic error}
\begin{equation}\begin{aligned}\label{eq:fqr}
  \fqr\err(A\|B) &\coloneqq \inf \lset \Tr (\id - M)^2 A + \Tr M^2 B \bar 0 \leq M \leq \id \rset.
\end{aligned}\end{equation}
This quantity admits some convenient representations. First, it can be thought of as a measured variant of a harmonic mean between $A$ and $B$, which recently made an appearance in~\cite{cheng_2026}. 
Equivalently, it measures the discrimination error of the pretty good measurement defined with respect to the Bures matrix quotient. Precisely, the standard pretty good measurement~\cite{hausladen_1994,barnum_2002} for distinguishing two positive operators $A$ and $B$ would consists of the POVM elements $\left( \frac{A}{A+B}, \frac{B}{A+B} \right)$, where $\frac{X}{Y} = Y^{-1/2} X Y^{-1/2}$. By replacing the matrix quotient $\frac{X}{Y}$ with $\J_{Y}^{-1}(X)$, we instead obtain the Bures pretty good measurement. The equivalence with fully quadratic error is formalised as follows.

\begin{boxed}
\begin{lemma}
It holds that
\begin{equation}\begin{aligned}
  \fqr\err(A\|B) = \Tr A - \norm{A}{A+B}^2 = \< A, B\>_{A+B}.
\end{aligned}\end{equation}
Furthermore, we have that $\fqr\err(A\|B) \leq \err(A\|B) \leq 2 \fqr\err(A\|B)$.
\end{lemma}
\end{boxed}
\begin{proof}
The first equality follows by strong Lagrange duality; the second is by rearranging the first, or by noting from the stationarity of the Lagrangian that the optimal measurement in~\eqref{eq:fqr} is $M = \J_{A+B}^{-1}(A)$. The inequality $\fqr\err(A\|B) \leq \err(A\|B)$ is then immediate from definition, as $M^2 \leq M$ and $(\id-M)^2\leq\id-M$. Taking $M = \J_{A+B}^{-1}(A)$ as a feasible solution for $\err(A\|B)$ gives $\err(A\|B) \leq \< A, B \>_{A+B} + \< B, A \>_{A+B} = 2 \fqr\err(A\|B)$.
\end{proof}

With this, the proof will follow a similar structure to~\cite[Theorem~1]{shen_2022}, with the key difference being the different matrix quotient employed here; this will allow us to state the bound in terms of the average error instead of R\'enyi entropies. 
We also make use of the fact that applying the pretty good measurement bin-by-bin only needs 2-universality, without requiring pairwise independence of the hash functions as in~\cite{shen_2022}.

\begin{boxed}
\begin{lemma}[{Converse of Theorem~\ref{thm:strongconverse}}] \label{lemma:strongconverse_converse}
For any 2-universal family of hash functions, it holds that
\begin{equation}\begin{aligned}
\EE_f \,\err\!\left( \rho_{ZE}^f  \middle\| \frac{\id_Z}{|\Z|} \otimes \rho_E \right) &\leq 4\, \fqr\err\!\left( \rho_{XE}  \middle\|  \frac{\id_X}{|\Z|} \otimes \rho_E \right)\\
&\leq 4 \err\!\left( \rho_{XE}  \middle\|  \frac{\id_X}{|\Z|} \otimes \rho_E \right).
\end{aligned}\end{equation}
\end{lemma}
\end{boxed}
\begin{proof}
Write $\rho^f_{E,z} \coloneqq \sum_{x: f(x) = z} \rho_{E,x}$, so that $\rho_{ZE}^f = \sum_z \proj{z} \otimes \rho^f_{E,z}$. Then
\begin{equation}\begin{aligned}
  \err\!\left( \rho_{ZE}^f  \middle\| \frac{\id_Z}{|\Z|} \otimes \rho_E \right) &= \sum_z \err\!\left( \rho^f_{E,z} \middle\| \frac{1}{|\Z|} \rho_E \right)\\
  &\leq 2 \sum_z \fqr\err\!\left( \rho^f_{E,z} \middle\| \frac{1}{|\Z|} \rho_E \right)\\
  &= 2 \sum_z \< \rho^f_{E,z},\, \frac{1}{|\Z|} \rho_E \>_{\rho^f_{E,z} + \frac{1}{|\Z|}\rho_E}\\
  &= 2 \sum_z \sum_{x: f(x) = z} \< \rho_{E,x}, \, \frac{1}{|\Z|} \rho_E \>_{\rho^f_{E,z} + \frac{1}{|\Z|}\rho_E}.
\end{aligned}\end{equation}
For each $x$, let us now construct an operator that represents all collision terms in the given bin:
\begin{equation}\begin{aligned}
  \rho^f_{E, \coll(x)} \coloneqq \sum_{\substack{x' \neq x:\\ f(x') = f(x)}} \rho_{E,x'},
\end{aligned}\end{equation}
so that  $\rho^f_{E,z} = \rho_{E,x} + \rho^f_{E,\coll(x)}$. We then continue as
\begin{equation}\begin{aligned}\label{eq:putting1}
  \err\!\left( \rho_{ZE}^f  \middle\| \frac{\id_Z}{|\Z|} \otimes \rho_E \right) &\leq 2 \sum_z \sum_{x: f(x) = z} \< \rho_{E,x}, \, \frac{1}{|\Z|} \rho_E \>_{\rho_{E,x} + \rho^f_{E,\coll(x)} + \frac{1}{|\Z|}\rho_E}\\
  &\leq 2 \sum_z \sum_{x: f(x) = z} \< \rho_{E,x}, \, \frac{1}{|\Z|} \rho_E + \rho^f_{E,\coll(x)} \>_{\rho_{E,x} + \rho^f_{E,\coll(x)} + \frac{1}{|\Z|}\rho_E}\\
  &= 2 \sum_{x} \fqr\err\!\left( \rho_{E,x} \middle\| \frac{1}{|\Z|} \rho_E + \rho^f_{E,\coll(x)} \right),
\end{aligned}\end{equation}
where in the second line we used that $\rho^f_{E,\coll(x)} \geq 0$ and that $\J_{A}^{-1}$ is a positive map for any positive semidefinite
$A$. Choosing $M_x$ to be an optimal measurement for $\fqr\err\!\left(\rho_{E,x} \middle\| \frac{1}{|\Z|}\rho_E\right)$, we can bound each term as
\begin{equation}\begin{aligned}\label{eq:putting2}
  \fqr\err\!\left( \rho_{E,x} \middle\| \frac{1}{|\Z|} \rho_E + \rho^f_{E,\coll(x)} \right) &\leq \Tr (\id - M_x)^2 \rho_{E,x} + \frac{1}{|\Z|} \Tr M_x^2 \rho_E + \Tr M_x^2 \rho^f_{E,\coll(x)}\\
  &= \fqr\err\!\left(\rho_{E,x} \middle\| \frac{1}{|\Z|}\rho_E\right) + \Tr M_x^2 \rho^f_{E,\coll(x)}.
\end{aligned}\end{equation}
Now, the 2-universality of $f$ gives
\begin{equation}\begin{aligned}\label{eq:putting3}
  \EE_f \, \Tr M_x^2 \rho^f_{E,\coll(x)} &=  \EE_f \sum_{\substack{x' \neq x}} \delta_{f(x),f(x')}\, \Tr M_x^2 \rho_{E,x'}\\
  &\leq \frac{1}{|\Z|} \sum_{\substack{x' \neq x}} \Tr M_x^2 \rho_{E,x'}\\
  &\leq \frac{1}{|\Z|} \sum_{\substack{x'}} \Tr M_x^2 \rho_{E,x'}\\
  &= \frac{1}{|\Z|} \Tr M_x^2 \rho_{E}\\
  &\leq \frac{1}{|\Z|} \Tr M_x^2 \rho_{E} + \Tr (\id - M_x)^2 \rho_{E,x}\\
  &=\fqr\err\!\left(\rho_{E,x} \middle\| \frac{1}{|\Z|}\rho_E\right) 
\end{aligned}\end{equation}
Putting~\eqref{eq:putting1}--\eqref{eq:putting3} together, we thus have
\begin{align}
  \EE_f \,\err\!\left( \rho_{ZE}^f  \middle\| \frac{\id_Z}{|\Z|} \otimes \rho_E \right) &\leq 4 \sum_x \fqr\err\!\left( \rho_{E,x} \middle\| \frac{1}{|\Z|} \rho_E \right) \nonumber\\
 &= 4\, \fqr\err\!\left( \rho_{XE}  \middle\|  \frac{\id_X}{|\Z|} \otimes \rho_E \right)\\
  &\leq 4 \err\!\left( \rho_{XE}  \middle\|  \frac{\id_X}{|\Z|} \otimes \rho_E \right).\tag*{\qedhere}
\end{align}
\end{proof}

\begin{remark}
The achievability result in Theorem~\ref{thm:strongconverse} can be compared with the classical, non-universal bound we derived earlier in Proposition~\ref{prop:nonuniv_collision_bound}, which can be understood as
\begin{equation}\begin{aligned}\label{eq:noside_halfquad}
  1 - \Delta_{\min}(p_X) \geq \hqr\err\left( p_X \middle\| \frac14 \frac{\id_X}{|\Z|} \right) \geq \frac34 \err\left( p_X \middle\| \frac14 \frac{\id_X}{|\Z|} \right).
\end{aligned}\end{equation}
Disregarding the constants, we see that this has the same exponential behaviour as the achievability in Theorem~\ref{thm:strongconverse}. However, the key difference is that there is no constant prefactor in the first inequality, which allows us to apply it as a bound on $\Delta_{\min}$, i.e.\ as an achievability bound on the \emph{error} exponent. It then leads to the tight exponential bound in Corollary~\ref{cor:collision_exp_noside}, since
\begin{equation}\begin{aligned}
 \lim_{n\to\infty} - \log \left[ 1 - \hqr\err\left( \rho_{XE}^{\otimes n} \middle\| \frac{\id_{X^n}}{\exp(nR)} \otimes \rho_E^{\otimes n} \right) \right] = \sup_{\alpha \in (1,2]} \,(\alpha-1) \left( \wt{H}^\downarrow_\alpha(X|E)_\rho - R \right).
\end{aligned}\end{equation}
But the inequality~\eqref{eq:noside_halfquad} is only known to hold for the case of no side information, so it does not look possible to use Theorem~\ref{thm:strongconverse} for achievability of the error exponent in the more general case. 

The difference can be seen from the interpolation functional form of the expressions that we encountered in our strong converse results compared to the ones we used in the study of error exponents: on the one hand, recalling the variational form of $\Ds{2}{\ve}{M}$, a rewriting of Lemma~\ref{lem:strongconv_lemma} gives 
\begin{equation}\begin{aligned}
  1 - \hqr\err(\rho\|\sigma) = \inf_{\ve\in[0,1)} \left[ \ve + \frac14 \exp\left( \Ds{2}{\ve}{M}(\rho\|\sigma) \right)\right],
\end{aligned}\end{equation}
while the half-quadratic hypothesis testing problem we encountered in Section~\ref{sec:universal} corresponds to
\begin{equation}\begin{aligned}
  \hqr\pi_{k} (\rho\|\sigma) = \inf_{\ve\in[0,1)} \left[ \ve + \sqrt{ k\, \exp\left( \Ds{2}{\ve}{M}(\rho\|\sigma) \right)}\right].
\end{aligned}\end{equation}
The key difference is the different power at which the collision entropy enters the optimisation, which strongly affects the asymptotic behaviour.
\end{remark}

\section{Discussion}

A conceptual contribution underlying our results here is a re-interpretation of the one-shot achievability results in terms of measured smooth R\'enyi entropies~\cite{regula_2026} in two different ways. First, we showed how they can be conveniently packaged as a framework based on quantum hypothesis testing problems that incorporate quadratic constraints, which can simplify the derivation of many properties and bounds. By extending this approach, we also established a new achievability result in the strong converse regime and proved its tightness. 
Second, we showed that the same optimisation exactly corresponds to a real interpolation functional between the standard $L_1$ operator space endowed with the trace norm and the $\smash{\rowcol}$ row--column space, which then naturally connected with proof techniques originating in operator space theory and non-commutative probability, ultimately making it possible for us to very sharply certify the optimality of the quantum leftover hash lemma of~\cite{regula_2026}.

While we do not believe that `half-quadratic' discrimination problems have been explicitly studied before, this is not the first occurrence of optimisation problems of a similar form. Expressions that in our notation correspond to $\smash{\hqr\pi}_k(\rho_A \| \id_A)$ already found use in the one-shot analysis of quantum data compression~\cite{abdelhadi_2020,farber_2026} and pure-state quantum resource distillation~\cite{regula_2017,regula_2019}. Interestingly, those operational tasks belong to a class of problems complementary to privacy amplification, namely, packing problems. This could suggest the potential for techniques based on half-quadratic hypothesis testing to find applications in quantum information more broadly, and we hope that the breadth of their applications that we demonstrated will help with this development.

In our discussions and derivations we have found it intuitively appealing to connect to hypothesis testing, but we also saw that the underlying structure was actually a case of real interpolation of operator spaces. This shows how the interpolation framework is fundamentally connected to how smooth entropies are applied in the one-shot analysis of quantum privacy amplification. 
Although \emph{complex} interpolation has found very important applications in the study of quantum information through connections with sandwiched R\'enyi divergences~\cite{beigi_2023}, and in particular enabled many of the tightest bounds on asymptotic exponents~\cite{wilde_2014,cheng_2022,dupuis_2023,cheng_2026-1}, real interpolation techniques have been somewhat underexplored in this context.
Our results suggest that real interpolation approaches can tell us even more about the precise behaviour of the error than the corresponding bounds from complex interpolation. This contrasts with the discussions in works such as~\cite{dupuis_2023,cheng_2026-1}, where complex interpolation was motivated as a way to overcome the limitations of smooth entropic bounds. 
Once again, since a very similar structure based on smooth entropies is shared by a variety of operational tasks in quantum information, we expect these ideas and techniques to find many further applications.

Our analysis here is particularly suited to the study of randomness extraction protocols based on $k$-universal hash functions, and all of our main results make use of assumptions of universality. 
As we discussed in detail in Section~\ref{sec:warmup}, already~\cite{hayashi_2013} showed that the behaviour of error exponents under trace distance can substantially change when non-universal (source-dependent) hashing is used. While such a relaxation may not reflect hashing protocols that would be employed in practice, at least from a mathematical perspective, it tells us that very different techniques would be needed in order to establish achievability and converse bounds that apply to all hash functions (cf.~\cite{regula_2026}). At the level of the underlying entropic quantities, monotonicity under all hash functions is reflected by a property called coarse graining~\cite{rubboli_review}; interestingly, while the sandwiched R\'enyi conditional entropies $\smash{\wt{H}_\alpha}$ do possess it, the Petz--R\'enyi conditional entropies $\smash{\ol{H}_\alpha}$ generally do not~\cite{goodarzi_2026}. For the particular case of strong converse exponents, combining our Theorem~\ref{thm:strongconverse} with~\cite[Proposition~26]{regula_2026} leads to a curious phenomenon where the exponent under all hash functions is bounded as
\begin{equation}\begin{aligned}
   \sup_{\alpha \in (0,1)} (\alpha-1) \left( \ol{H}^{\downarrow}_{\alpha}(X|E)_\rho - R \right) &\geq \lim_{n\to\infty} - \frac1n \log \Big[ 1 - \Delta_{\min}\!\left(\rho_{XE}^{\otimes n},\, \exp(nR) \right)\Big] \\&\geq \sup_{\alpha \in (0,1)} (\alpha-1) \left( \wt{H}^{\downarrow}_{\alpha}(X|E)_\rho - R \right),
\end{aligned}\end{equation}
but neither of the bounds can be expected to be tight in general.  
This could hint that other conditional entropies, such as the more general families of~\cite{rubboli_2026-3}, may be needed to fully characterise non-universal quantum privacy amplification. Indeed, recent developments showed the usefulness of such conditional entropies in characterising variants of randomness extraction~\cite{li_2026-2,rubboli_2026} and related problems like quantum state merging~\cite{berta_2026-1}, albeit under different distance measures. 

Furthermore, while the strong converse results of Section~\ref{sec:strong-converse} apply broadly to 2-universal hashes, our tight converse bound on the error in Section~\ref{subsec:tight_random_binning} makes use of 4*-universality and in general may not apply if this condition is relaxed. The appearance of fourth-moment assumptions in this context is not surprising: as in the fourth-moment method~\cite{berger_1997} that finds use in the analysis of algorithms and information-theoretic problems, estimates of fourth moments are generally needed to give anticoncentration-type lower bounds on the expected absolute value of a random variable, such as when lower bounding the expected trace norm. While 4-universal hash functions are much less idealised than the fully universal random-binning constructions used in the converse bounds of~\cite{hayashi_2013,hayashi_2016-3,li_2026-1}, and indeed efficient implementations of 4-universal hash functions do exist~\cite{thorup_2004}, one general interpretation of this is an indication of the tightness of the leftover hash lemma formulated in~\cite{regula_2026} for any quantum state; for other families of hash functions and for specific types of source states, more accurate bounds may certainly exist.

Finally, let us comment on the fact that our results apply to the specific choice of the error metric being defined in terms of the trace distance. 
We consider this as one of the key technical strengths of our approach, as it is the trace distance that is needed for cryptographic applications~\cite{portmann_2022} where the leftover hash lemma is most prominently used. Indeed, significant improvements in the analysis of quantum key distribution were made possible by the tight achievability results that we studied here~\cite{tan_2026}, but the optimality of such bounds in the sense of asymptotic scaling is only established in this work. 
Nevertheless, in other contexts --- such as when going beyond classical randomness to fully quantum tasks like decoupling~\cite{dupuis_2014,dupuis_2010} --- estimates of other distance measures such as the fidelity may be needed. The behaviour of the operational problems like randomness extraction is very different under other error criteria, as are the techniques needed to study them~\cite{hayashi_2015,li_2023,li_2024,cheng_2025-1,rubboli_2026,li_2026}. Understanding whether the technical developments of this work can shed light also on those other settings presents an exciting avenue for further research.


\medskip
\textbf{Acknowledgments.} --- We thank Roberto Rubboli and Marco Tomamichel for many helpful discussions and comments. 
B.R. acknowledges the support of the Japan Science and Technology Agency (JST) PRESTO grant no.\ JPMJPR25FB.
H.C.~acknowledges the support by NSTC 115-2628-E-002-005, NSTC 114-2119-M-001-002, and NSTC 115-2124-M-002-014, NTU-115V2016-1, NTU-CC115L893705, and the Yushan Fellow Program by the Ministry of Education (MOE) MOE-114-YSFEE-0003-006-P2.

\medskip
\textbf{Statement on use of AI.} --- All content in the paper itself is of biological origin. However, we benefited from discussions with ChatGPT 5.5 and 5.6 Sol at many points of our research. The LLMs have in particular assisted with studying the relevant literature on interpolation functionals in probability theory and developing the proof approach in Section~\ref{subsec:tight_random_binning}, as well as with streamlining and optimising parts of other proofs conceived by the human authors; also, the simple collision argument we used to show Proposition~\ref{prop:nonuniv_collision_bound} is due to ChatGPT. 

\medskip
\textbf{Note added.} --- See also the concurrent work~\cite{concurrent} by Hao-Chung Cheng, Li Gao, Shi-Bing Li, and Hongsen Qiu for related results on the trace-distance exponents for privacy amplification.  


\excludetoc

 \bibliographystyle{apsca}
 \bibliography{main}


\end{document}